\documentclass[aps,reprint,groupedaddress,showpacs]{revtex4-1}
\usepackage{amsmath}
\usepackage{amssymb}
\usepackage{color}
\usepackage[colorlinks,bookmarks=false,citecolor=red,linkcolor=blue,urlcolor=blue]{hyperref}
\usepackage{graphicx}
\usepackage[caption=false]{subfig} 
\usepackage{wrapfig}

\def\Hhat{\hat{H}}

\def\nhat{\hat{n}}
\def\chat{\hat{c}}

\def\psihat{\hat{\psi}}
\def\phihat{\hat{\phi}}
\def\Qhat{\hat{Q}}

\begin{document}
\title{Edge-modes in a frustrated quantum Ising chain}
\author{Somenath Jalal}
\email{somenath.jalal@gmail.com}
\author{Brijesh Kumar}
\email{bkumar@mail.jnu.ac.in}
\affiliation{School of Physical Sciences, Jawaharlal Nehru University, New Delhi 110067, India}
\date{\today}


\begin{abstract}
The ground state properties of an Ising chain with nearest ($J_{1}$) and next-nearest neighbor ($J_{2}$) interactions in a transverse field are investigated using the density matrix renormalization group and cluster mean-field theory methods. Its quantum phase diagram has two ordered regions (separated by a quantum disordered phase), one of which has simple ferromagnetic or N\'eel order depending upon the sign of $J_{1}$, and the other is the double-staggered antiferromagnetic phase. The presence of the Majorana-like edge-modes in these ordered phases is inferred by calculating the end-to-end spin-spin correlations on open chains. It is found that there occur four edge modes in the double-staggered phase, while the ferromagnetic and N\'eel phases support two edge modes, except very near $J_{1}=0$ where it seems they have four edge modes like the exact case at $J_{1}=0$ itself.
\end{abstract}
\maketitle

\section{\label{sec:intro} Introduction}
The Ising model in a transverse field is an important physical problem.  It is also referred to as the quantum Ising (QI) model because the presence of a transverse field causes quantum fluctuations to  Ising spins. Historically, the interest in this model started through a work of de Gennes where it was used to model the proton dynamics in the hydrogen-bonded ferroelectric materials ($e.g.$, KH$_{2}$PO$_{4}$)~\cite{deGennes}. The QI model continues to be a subject of diverse current interests, as clear from a variety of physical contexts in which it occurs~\cite{book.Bikas, randomQI.Fisher, Moessner, QI.Holo.Girvin, Coldea, bermudez-qi-expt,Mila.Square}. A notable recent case, for instance, is that of the Rabi lattice model which, in the strong coupling limit exhibits, QI dynamics~\cite{bkumar.somenath}.

The one-dimensional (1D) spin-1/2 QI model with only nearest-neighbor interaction, $J_1$, is exactly soluble under Jordan-Wigner (JW) fermionization~\cite{pfeuty}. It undergoes a quantum phase transition from a disordered to a doubly degenerate ordered ground state as the strength of $J_1$ increases relative to the transverse field, $h$. In the fermionic form, it becomes Kitaev's superconducting quantum wire that harbours two Majorana fermion modes at the free ends of an open chain~\cite{Kitaev.QWire}. These Majorana-like edge-modes arise only in the ordered phase, and do not exist in the disordered phase. There is much current interest in realizing the Majorana modes for quantum computation~\cite{Nayak.RMP, jason-majorana}, and a possible way of achieving this could be through Kitaev's quantum wire, viz., the 1D QI model. 

The Majorana edge modes in Kitaev's quantum wire are topologically protected, as no local perturbation can couple these modes sitting at two opposite ends of the wire. But in the QI chain where the basic physical variables are spins, not fermions, the longitudinal field acts unfavourably upon them. However, through energetic (if not topological) considerations, the edge-modes in a QI chain still stand a chance of survival against such detrimental perturbations, as discussed recently by the present authors~\cite{bkumar.somenath}. There, we used the relation, $\rho^x_{1L} = p^8$ (derived by Pfeuty in Ref.~\onlinecite{pfeuty}), between the end-to-end spin-spin correlation, $\rho^x_{1L}$, and the order parameter, $p$,  as a  signature of these edge-modes, and found that it is satisfied even in the presence of the longitudinal fields (uniform as well as random) for strong enough $J_1$ under suitable conditions. Continuing with our studies of the edge modes in the presence of longitudinal operators, we further like to understand the effects of the Ising interactions beyond nearest-neighbour. 

Our basic motivation is to investigate the occurrence of the edge modes in a frustrated quantum Ising problem. Therefore, in this paper, we study the $J_1$-$J_2$ QI model on an open chain. As described in Eq.~(\ref{eq:model}) of Sec.~\ref{sec:model}, this model has a next-nearest neighbour Ising interaction, $J_{2}$, in addition to $J_{1}$ and $h$. It is a minimal QI problem that has frustration. It is also known in the literature as quantum ANNNI (anisotropic or axial next-nearest neighbour Ising) model, because it is the quantum equivalent of the two-dimensional classical ANNNI model (through the transfer matrix in statistical mechanics)~\cite{QIj1j2.Barber,ANNNI.Selke}. Although it is a well-studied problem, the edge modes, it seems, have never been investigated in this model. Since this problem is not exactly soluble, we study it numerically by employing DMRG (density matrix renormalisation group) and cluster-mean-field theory (CMFT) methods. The calculations and their results are discussed in Secs.~\ref{subsec:QPD} and~\ref{sec:EM}. Through these calculations, we particularly look for the signatures of the edge-modes in the ordered phases of the $J_{1}$-$J_{2}$ QI chain by computing the end-to-end spin-spin correlations. We find that the regions dominated by $J_{2}$ has four edge-modes, while the rest of it support two edge-modes. The main results are summarised in Fig.~\ref{fig:summary} in Sec.~\ref{sec:sum}.

\section{\label{sec:model} The $J_1$-$J_2$ quantum Ising model}
\begin{figure}[t] 
\centering
\includegraphics[width=0.4\textwidth]{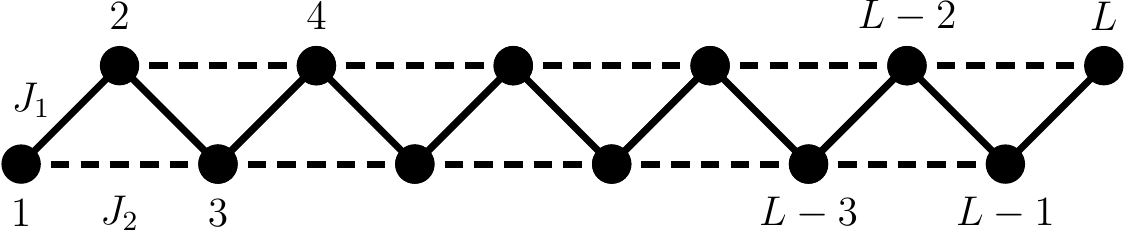} 
\caption{The $J_1$-$J_2 $ quantum Ising chain of $L$ spins. The solid lines denote $J_{1}$ interaction, and the dashed lines $J_{2}$.}
\label{fig:model}
\end{figure}

On a chain with open boundaries, the Ising model with nearest and next-nearest neighbour interactions in a transverse field can be written as follows.
\begin{eqnarray}
\Hhat = J_1 {\sum_{i=1}^{L-1}} {\sigma^x_{i} \sigma^x_{i+1}} + J_2 {\sum_{i=1}^{{L-2}}} {\sigma^x_i \sigma^x_{i+2}} + h {\sum_{i=1}^{L}} {\sigma^{z}_{i}} {\label{eq:model}} 
\end{eqnarray}
Here, $\sigma_i^x$ and $\sigma_i^z$ are the Pauli operators, and $L$ is the total number of spins. It is also depicted in Fig.~\ref{fig:model}.

\begin{figure}[b]
\centering
\includegraphics[width=0.35\textwidth]{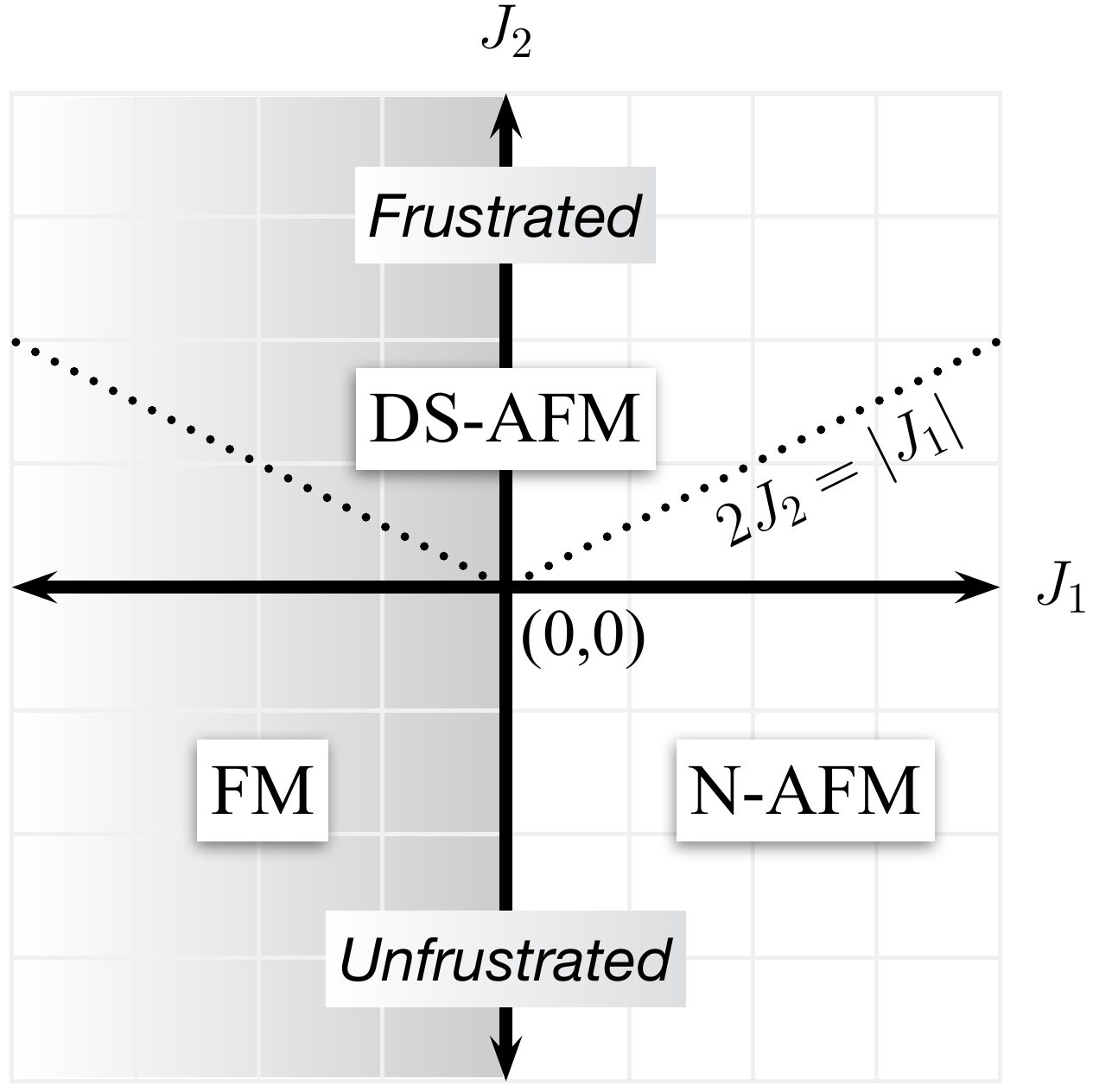} 
\caption{The frustrated and unfrustrated regions of $\Hhat$ in the $J_1$-$J_2$ plane. The dotted line, $2J_{2}=|J_{1}|$, is the line of maximum frustration, which together with $J_{1}=0$ line for negative $J_{2}$ form the phase boundaries between three phases, FM, N-AFM and DS-AFM, of the classical $J_{1}$-$J_{2}$ Ising chain.
}
\label{fig:j1j2plane}
\end{figure}

The $J_1$ and $J_2$ terms in $\Hhat$ compete for setting the order in the ground state. Whether the competing interactions frustrate the spins, or not, is decided by their signs and relative magnitudes. Of the four quadrants in the $J_1$-$J_2$ plane, as shown in Fig.~\ref{fig:j1j2plane}, the two in the lower half-plane for $J_2<0$ correspond to the unfrustrated cases. This is because a negative $J_{2}$ does not act against $J_{1}$. For instance, a positive $J_{1}$ favours N\'eel antiferromagnetic (N-AFM) order in which the nearest neighbour spins are anti-parallel, and therefore, the second neighbour spins are parallel to each other. This simultaneously satisfies a negative $J_{2}$ and a positive $J_{1}$. Hence, no frustration.  Likewise for $J_{1}<0$, which favours ferromagnetic (FM) order. On the other hand, a positive $J_{2}$, which favours the anti-parallel alignment of the second-neighbor spins, always competes against the $J_{1}$ of either sign. Hence, $J_2>0$ is the frustrated case. 

A property of the $\Hhat$ of Eq~(\ref{eq:model}) is that $J_{1}$ exactly maps to $-J_{1}$, without affecting $J_{2}$ and $h$, under the transformation $\{\sigma^{x}_{i}\} \rightarrow -\{\sigma^{x}_{i}\}$ on any one of the two sub-lattices (even or odd sub-chain in Fig.~\ref{fig:model}). It would suffice,  therefore, to study $\Hhat$ either for negative or positive $J_{1}$ only, as the physics of one case is an image of the other. For instance, the FM order becomes N-AFM under this mapping. In this paper, we do calculations only for $J_{1}<0$, the shaded left-half of the $J_{1}$-$J_{2}$ plane in 
Fig.~\ref{fig:j1j2plane}. 

While the FM or N-AFM ordering in the ground state is supported naturally by $J_{2}<0$, these phases  also extend energetically into the frustrated upper-half plane. For the classical case, that is $h=0$, one finds that the FM/N-AF phases extend up to $J_{2} < |J_{1}|/2$. At precisely, $J_{2}=|J_{1}|/2$, the dotted MG (Majumdar-Ghosh~\footnote{The Majumdar-Ghosh model is a historic 1D problem of frustrated quantum (Heisenberg) spin-1/2's~\cite{MG}. Since it precisely corresponds to $J_2=J_1/2$, it has become common to refer to this relation between $J_1$ and $J_2$ as MG point or line (depending upon the context).}) line in Fig.~\ref{fig:j1j2plane}, the frustration is maximum, with macroscopic degeneracy and no unique order in the classical ground state. Above this line, the $J_{2}$ dominates, and the classical ground state exhibits double-staggered antiferromagnetic (DS-AFM) order, that is, $|\cdots ++--++-- \cdots\rangle$~\footnote{The double-staggered AFM here is same as the `antiphase' in the conventional ANNNI literature.}. Here, $|+\rangle$ and $|-\rangle$ are the eigenstates of $\sigma^{x}$, such that $\sigma^{x}|\pm\rangle = \pm|\pm\rangle$.

The transverse field, assisted by strong frustration around the MG line, is expected to produce an extended quantum disordered region on both sides of the MG line. But sufficiently away from the MG line, the Ising interactions would overcome the transverse field to generate the expected classical orders. Below, we present the quantum phase diagram of the $J_{1}$-$J_{2}$ QI chain. Although it has been variously studied for its quantum phase 
diagram~\cite{QIj1j2.Barber,QIj1j2.Bikas,QIj1j2.Brazil,QIj1j2.Subinay,QIj1j2.DMRG,QIj1j2.Adam,review.dutta}, we compute it anyway because we will need it later for our investigations of the edge modes. Besides, we plot it differently.  As in Fig.~\ref{fig:j1j2plane}, we treat $J_{1}$ and $J_{2}$ as free parameters, and put $h=1$ for the rest of the discussion. Since $\Hhat$ is not amenable to exact solutions, except for $J_{2}$ or $J_{1}=0$, 
the calculations presented here are numerical in nature. But through them, we gain a fair understanding of the ground state properties of $\Hhat$.

\subsection{\label{subsec:QPD} Quantum Phase Diagram} 
To generate the quantum phase diagram of $\Hhat$, we compute spin-spin correlation, energy-gap, order parameters and transverse polarisation in the ground state of $\Hhat$. We do calculations only for $J_{1}<0$, as it contains complete information about positive $J_{1}$. Since DMRG is a nice method for studying 1D quantum systems, we too use it here for our problem. It is a numerical method that iteratively truncates the Hilbert space by keeping only the most probable contributions to the ground state, and thus allows access to large system sizes~\cite{dmrg1,dmrg2}. We also do cluster-MFT calculations on small chains.  

\begin{figure}[t] 
   \centering
   \includegraphics[width=.4\textwidth]{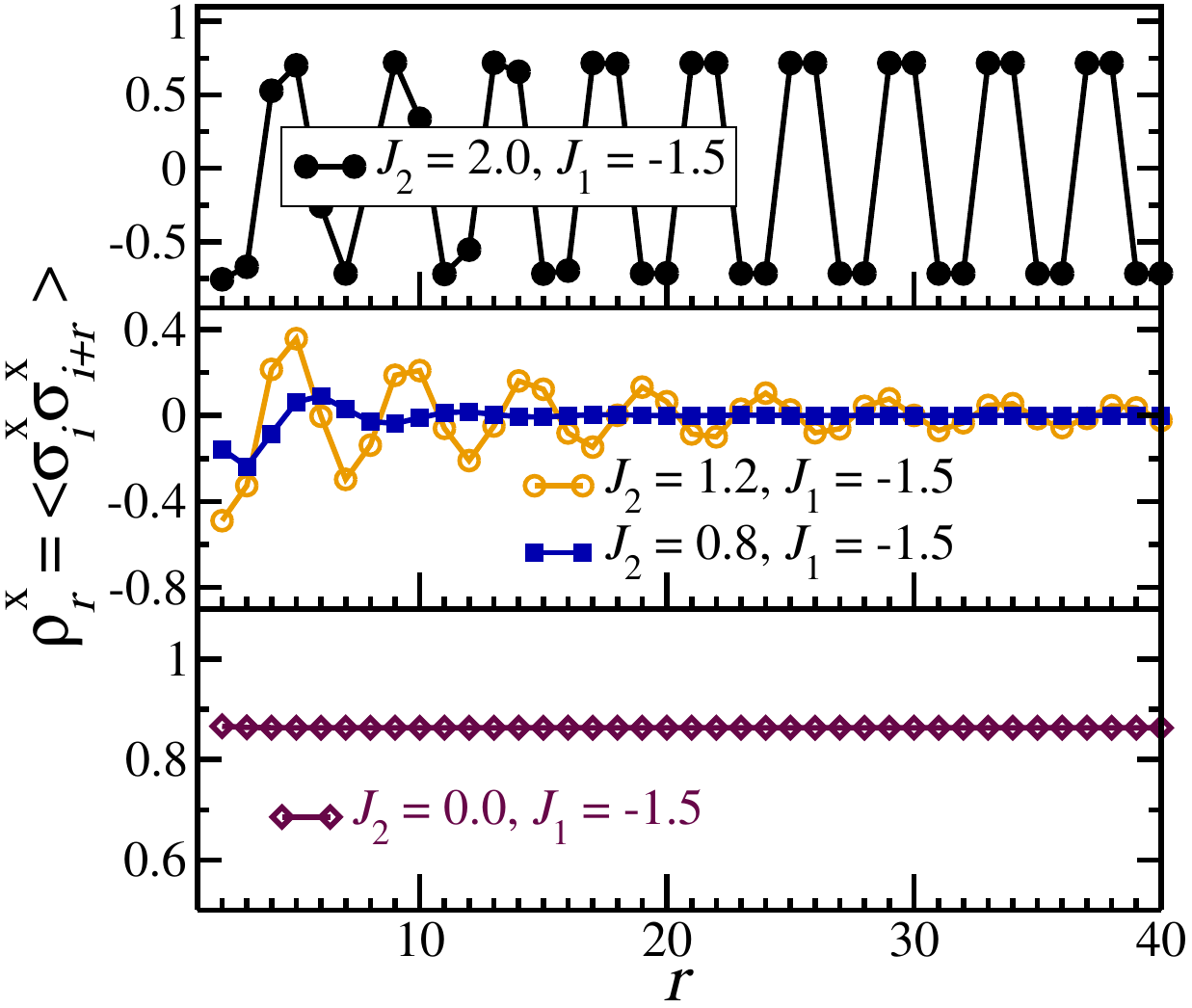} 
   \caption{The spin correlation functions in different phases.}
   \label{fig:xx-corr}
\end{figure}

The spin correlation function, $\rho^{x}_{r} = \langle \sigma^{x}_{i}\sigma^{x}_{i+r}\rangle$, in the ground state shows three different types of behaviour for different interaction strengths, as plotted in Fig.~\ref{fig:xx-corr}. For a fixed $J_{1}$, we calculate $\rho^{x}_{r}$ for $J_{2}$ sufficiently below and above the MG line, and in-between around it. Expectedly, the ground state of $\Hhat$ shows long-ranged FM and double-staggered AFM correlations for $J_{2}$ well below and above the MG line, respectively. Near MG line, on both sides, where the transverse field helped by frustration has the best chance to kill the order, we find that $\rho^{x}_{r}$ decays to zero in two noticeably different ways. Look at the middle panel in Fig.~\ref{fig:xx-corr} carefully. For a $J_{2}$ a bit farther above the MG line, $\rho^{x}_{r}$ decays visibly slowly compared to elsewhere in the disordered region, where it decays to zero very rapidly. This is consistent with the known algebraic and exponential decay behaviours~\cite{QIj1j2.DMRG,QIj1j2.Adam}.  

\begin{figure}[t] 
   \centering
   \includegraphics[width=.42\textwidth]{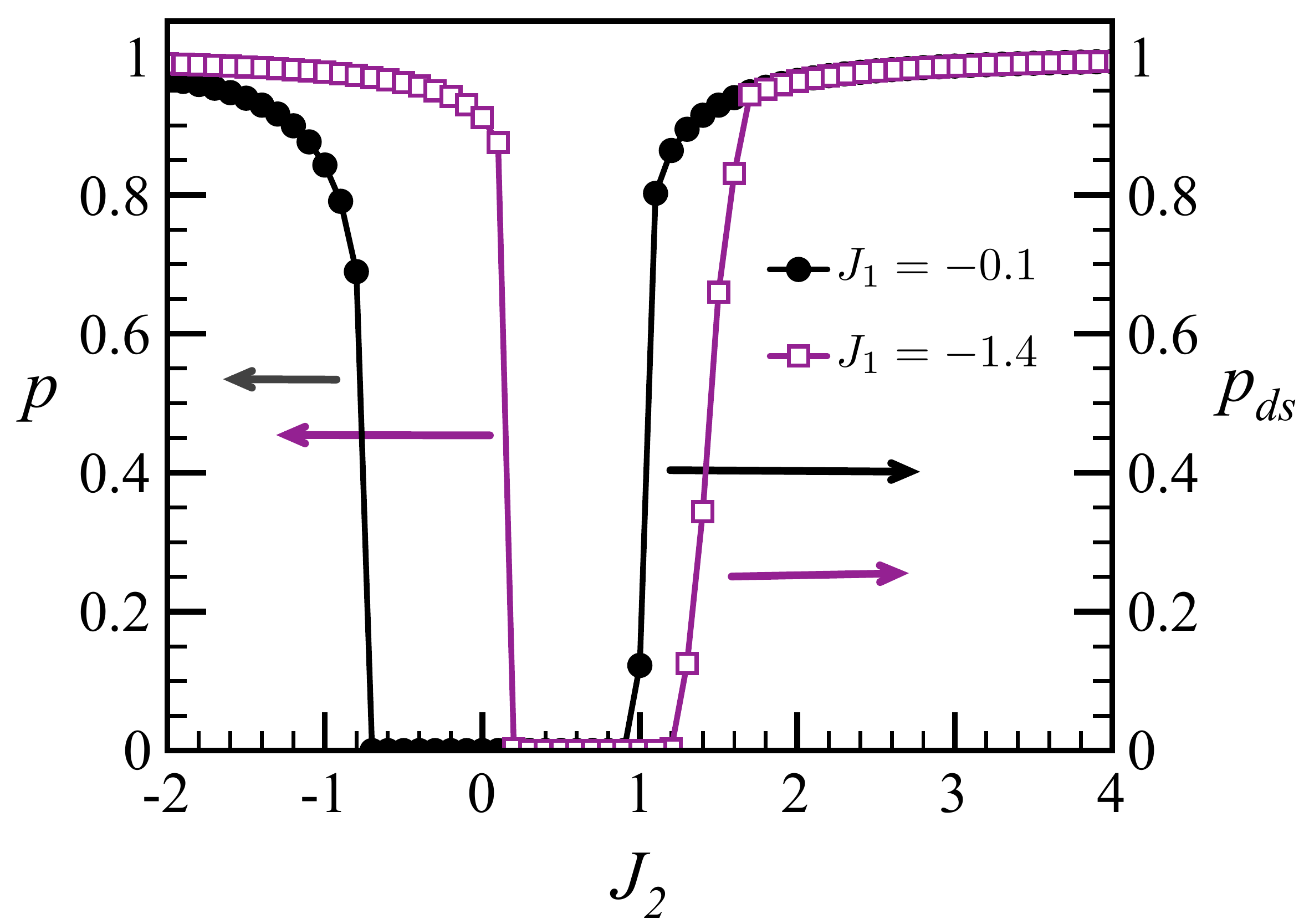}
   \caption{The order parameters vs. $J_{2}$ for fixed values of $J_{1}$.}
   \label{fig:p-pds}
\end{figure}

The FM order parameter, $p$, is the ground state expectation of the uniform spin polarisation, $p=\frac{1}{L}\sum_{i}\langle\sigma^{x}_{i}\rangle$. Moreover, the long-range FM order implies $\rho^{x}_{r\rightarrow\infty} = p^{2}$. We use these two definitions to compute $p$,  both of which give consistent results. In the DS-AFM phase, the order parameter, $p^{}_{ds}$, is defined as $ p_{ds}^{2}=|\rho^{x}_{r \rightarrow\infty} | $, or as $p^{ }_{ds}=\frac{1}{L}\sum_{n=1}^{L/2}(-)^{n}\langle\sigma^{x}_{2n-1} +\sigma^{x}_{2n}\rangle$. Figure~\ref{fig:p-pds} presents the variation of $p$ and $p^{ }_{ds}$ along the $J_{2}$ axis for fixed values of $J_{1}$.  It reveals two quantum phase transitions, separately characterised by the vanishing of $p$ and $p_{ds}$.
 
\begin{figure}[t] 
   \centering
   \includegraphics[width=.4\textwidth]{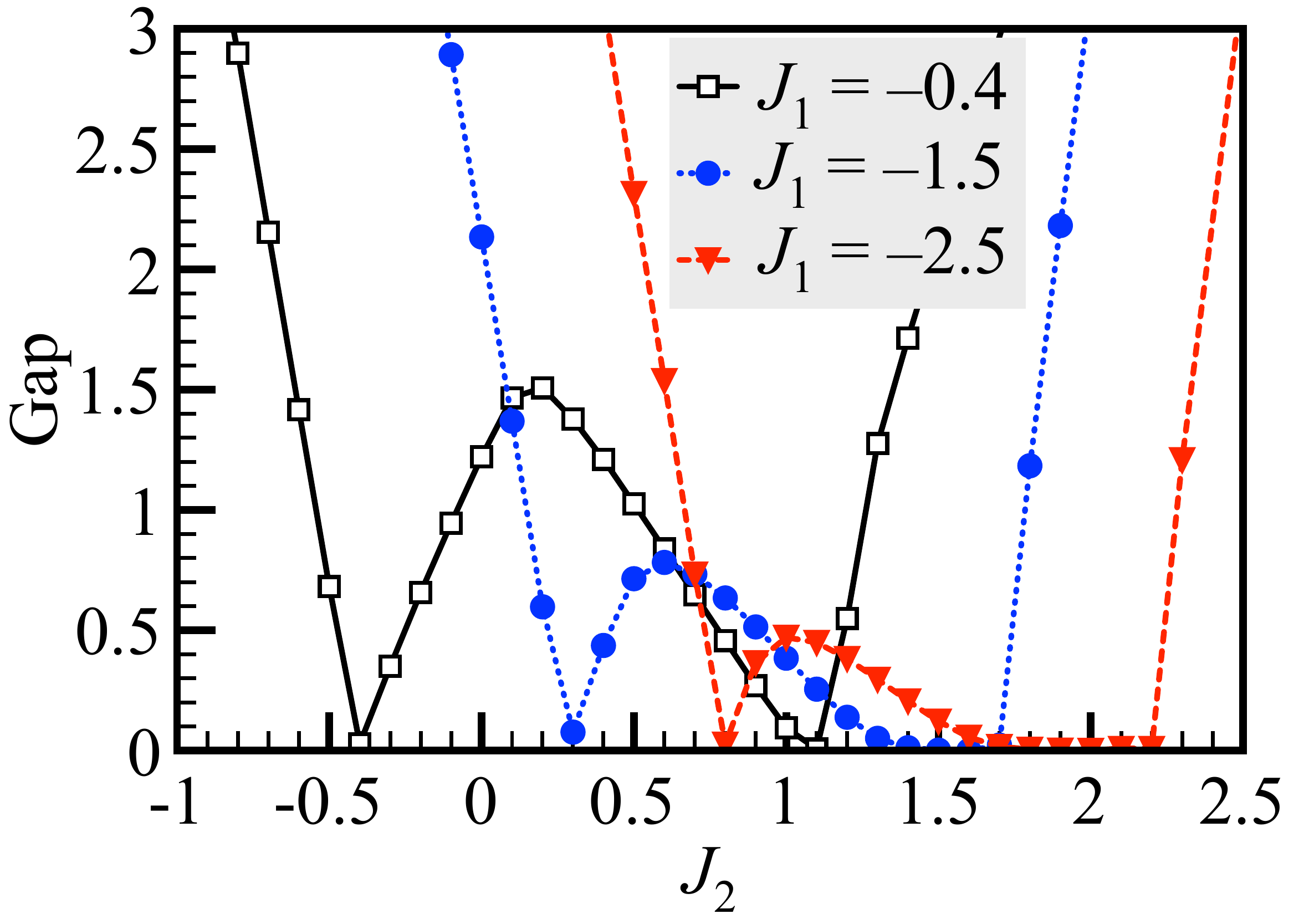}
   \caption{Energy-gap from DMRG calculation for $L=400$.} 
   \label{fig:gap}
\end{figure}

As $\Hhat$ only has a discrete (parity) symmetry, we find both the ordered phases to be gapped, as shown in Fig.~\ref{fig:gap}. The two gaps are zero at the respective critical points, and grow continuously to non-zero values in the ordered phases. On the disordered side of the FM critical point, the gap is again non-zero and continuously varying. It is the same region of parameters in which $\rho^{x}_{r}$ decays exponentially. We term this phase as quantum paramagnetic (Q-PM). The quantum disordered state adjoining the DS-AFM phase shows gaplessness in a small range of $J_{2}$ for a fixed $J_{1}$ (beyond which it becomes the gapped Q-PM phase). This is as if the critical point has expanded into a finite region of criticality. This gapless critical phase is the one in which one finds the algebraic spin-spin correlation. We call it a critical quantum paramagnetic (cQ-PM) phase~\footnote{In the ANNNI literature, this phase is called the `floating' phase, due to the associated physical picture in the classical context}. 

\begin{figure}[t] 
   \centering
   \includegraphics[width=.35\textwidth]{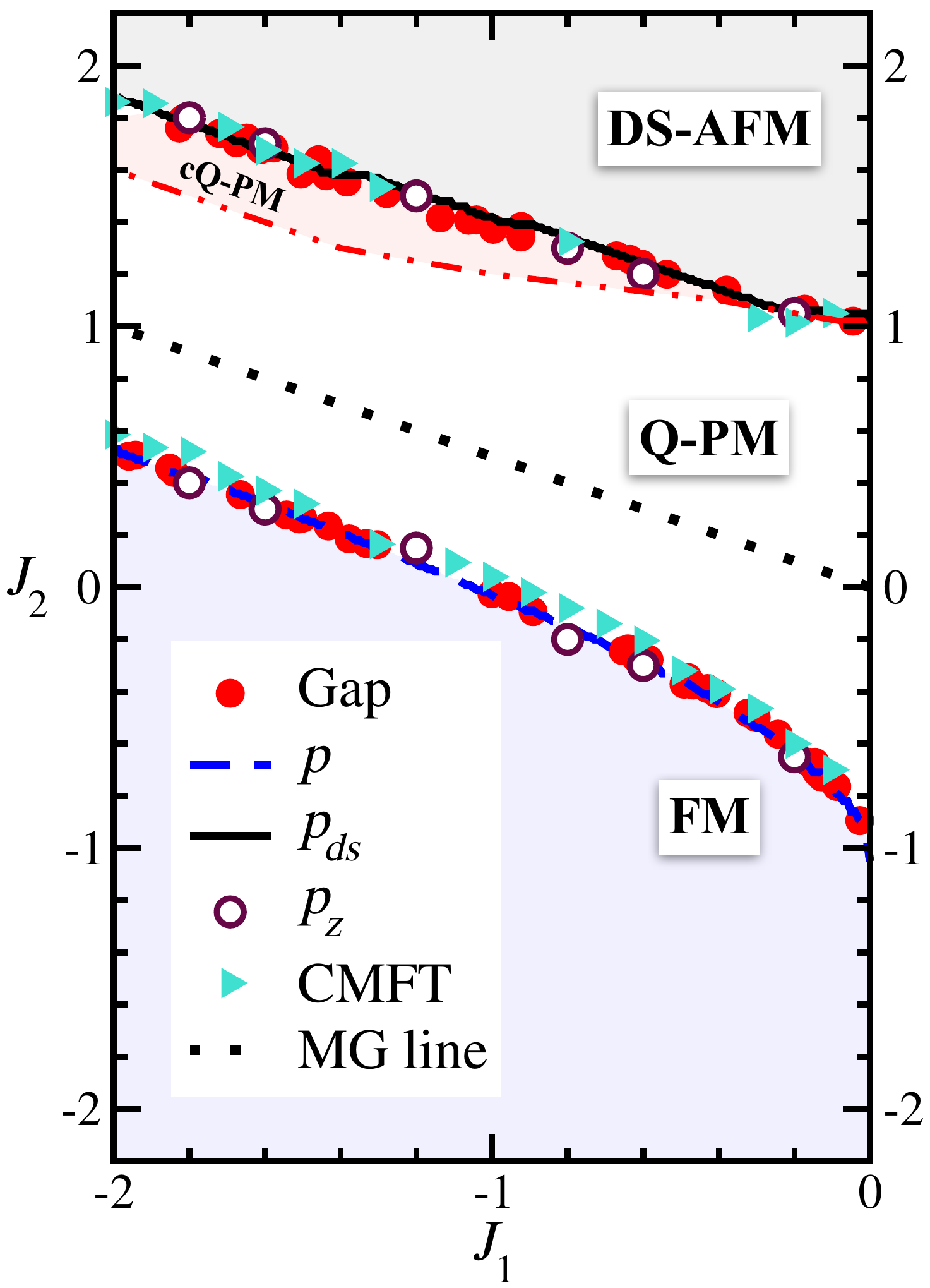}
  \caption{Quantum phase diagram of the one-dimensional $J_{1}$-$J_{2}$ quantum Ising model from DMRG and CMFT calculations.}
   \label{fig:QPD}
\end{figure}

By scanning through the energy-gaps and the order parameters in the $J_{1}$-$J_{2}$ plane, we identify the regions of FM, Q-PM, cQ-PM and DS-AFM phases. The resulting quantum phase diagram of $\Hhat$ is given in Fig.~\ref{fig:QPD}. Its basic topology is similar to the corresponding classical case shown in Fig.~\ref{fig:j1j2plane}, except that the disordered MG line has now turned into a quantum disordered region (consisting of Q-PM and cQ-PM phases) bounded by the critical lines of quantum phase transition to DS-AFM phase on the upper side and to FM phase on the lower side. The phase boundaries generated by tracking the gaps and the order parameters, from DMRG calculations on the chains of lengths up to 600, are consistent with each other. We also compute, $p_{z}=\frac{1}{L}\sum_{i=1}^{L}\langle\sigma^{z}_{i}\rangle$, the transverse polarisation in the ground state. For the exactly soluble QI problem, $p_{z}$ is known to have a kink at the critical point, while varying continuously across it. Likewise, we also find the kink-points of $p_{z}$ to fall on the critical lines found from the gap and order parameter calculations. 

The CMFT data in Fig.~\ref{fig:QPD} is $1/L \rightarrow 0$ extrapolation of the finite $L$ critical points from the cluster-mean-field calculations on the chains of length 4, 8 and 12. We numerically diagonalise $\Hhat$ coupled to a mean-field order parameter through boundary spins, and determine the order parameter self-consistently. The CMFT models we studied are: $ \Hhat +  p[(J_1 + J_2)(\sigma^{x}_{1}+\sigma^x_L) +  J_2(\sigma^{x}_{2}+\sigma^x_{L-1})]$ for the FM phase, and $ \Hhat +  p^{ }_{ds}[(J_1 + J_2)(\sigma^{x}_{1}-\sigma^x_L) +  J_2(\sigma^{x}_{2}-\sigma^x_{L-1})]$ for the DS-AFM phase.
The critical points thus calculated also agree with the DMRG data. 

These numerical phase boundaries correctly start from the exact critical points, $(J_{1},J_{2})=(0,\pm 1$), of two decoupled $J_{2}$-only QI chains. Moreover, the lower one correctly goes through the exact critical point $(-1,0)$. They also become parallel to the MG line for large $J_{1}$ and $J_{2}$.

\section{\label{sec:EM} Edge modes in the ordered phases}
The nearest-neighbour QI model on an open chain is famously known to have two exact Majorana edge modes in the ordered phase~\cite{Kitaev.QWire}. To see this, apply Jordan-Wigner transformation on $\Hhat$, and look at the trivial case of $J_{2}=h=0$. The JW transformation is a canonical and invertible map that relates the Pauli operators to the spinless fermions. It can be defined as  $\sigma^{+}_{i} = {\chat^{\dagger}_{i}} \prod_{l=1}^{i-1}\Qhat_{l}$ and $\sigma^{z}_{l}  = - \Qhat_{l} $, where,  $\chat^{\dag}_{l}$'s are the fermion creation operators, $\Qhat_{l}=e^{i \pi \nhat_{l}}$, and $\nhat_{l}=\chat^{\dag}_{l}\chat^{ }_{l}$. Moreover, $\chat^{\dag}_{l} = \frac{1}{2}(\phihat_{l} + i\psihat_{l})$, where $\phihat_{l}$ and $\psihat_{l}$ are two Majorana (Hermitian) fermions  that anticommute mutually and with other fermions, and $\phihat^{2}_{l}=\psihat^{2}_{l}=1$. Under this transformation, the $\Hhat$ of Eq.~(\ref{eq:model}) takes the following form.
 \begin{equation}
 \Hhat_{1} = J_{1} {\sum_{l=1}^{L-1}}  i \psihat_{l} \phihat_{l+1} + J_{2} {\sum_{l=1}^{L-2}}  i \psihat_{l} \Qhat_{l+1} \phihat_{l+2} + h {\sum_{l=1}^{L}} i \psihat_{l} \phihat_{l}
 \label{eq:model-JW}
 \end{equation}

In the simplest case of $J_{2}=h=0$, the Majorana operators, $\phihat_{1}$ and $\psihat_{L}$, at the ends of the chain do not figure in $\Hhat_{1}$. Thus, it has a zero energy eigen-mode described by two Majorana fermions localised at the opposite edges. Even when $h\neq 0$, but $J_{2}=0$, this problem can be solved exactly. This is what Pfeuty and Kitaev did, and found that in the ordered phase these edge-modes occur with an amplitude that is not strictly localised at the edges but decays with a finite spread into the bulk. However, when $J_{2}$ is also non-zero, things become difficult. A simplified version of this problem, in which  $i \psihat_{l} \Qhat_{l+1} \phihat_{l+2}$ in the $J_{2}$ term is replaced by  $i \psihat_{l} \phihat_{l+2}$ (by dropping $\Qhat_{l+1}$), has been given some attention recently, because it is a bilinear fermion problem amenable to exact solutions~\cite{niu2012prb}. But this is  not same as studying $\Hhat$. 
Since it is not quite studied what happens to the edge modes in the $J_{1}$-$J_{2}$ QI chain, we attempt to answer it here in this section. 

Consider the case of $h=0$, and $J_{1}$ and $J_{2}\neq 0$. Here too, $\phihat_{1}$ and $\psihat_{L}$ are absent in $\Hhat_{1}$. That is, the edge modes do exist for any $J_{2}$, at least in the absence of the transverse field. In fact, by this observation, they would exist even for longer range Ising interactions, as long as $h=0$. This clearly suggests that two Majorana-like edge modes may occur in the ground state of the $J_{1}$-$J_{2}$ model even when $h\neq 0$. Another case to note is that of $J_{1}=0$, and $J_{2}$ and $h$ non-zero. In this case, we have a problem of two independent exactly solvable QI chains (see Fig.~\ref{fig:model}), which realise `four' edge modes, two for each sub-chain. Encouraged by these observations, we now look for the edge-modes in the full $J_{1}$-$J_{2}$-$h$ problem. With no immediate help from analytics, we focus on numerics using DMRG and CMFT. It turns that our simple-minded CMFT calculations also prove quite helpful in a clear analysis of the edge modes.

Since the basic variables in $\Hhat$ are spins, not fermions, we need to devise suitable means to infer the presence of Majorana-like edge modes directly in terms of spins. In this context, Pfeuty's relation, $\rho^{x}_{1L} = \langle \sigma^{x}_{1} \sigma^{x}_{L} \rangle = p^{8}$, in the ordered phase of the nearest-neighbor QI chain with open boundaries, becomes particularly important to us. 
Just contrast it with the correlation between any two far-away spins in the bulk behaving as $p^{2}$.
This $p^{8}$ behaviour of the end-to-end spin-spin correlation is an exact indicator of the Majorana-like edge modes in the nearest neighbour QI chain~\cite{pfeuty}.  While no such relation is known for the $J_{1}$-$J_{2}$ problem, we still like to use it {\em empirically} as a signature of the edge-modes. Our past experience has been encouraging in this regard, as we have used this relation to draw meaningful inferences on the edge-modes in the 1D Rabi lattice, and in the QI chain with longitudinal field~\cite{bkumar.somenath}.

We set two simple rules for the analysis of the edge modes. The first rule states that `if any long-ranged spin-spin correlation goes as $p^{2}$ (or $p_{ds}^{2}$), it can only be a bulk correlation'. That is, for the edges to be special, their spin-spin correlation {\em must not} behave as $p^{2}$. The second rule is that `if the correlation between the spins on the edges approaches $p^{8}$ (or $p_{ds}^{8}$), as one goes deeper into the ordered phase, then it confirms the occurrence of the edge modes therein'. Our second rule is obviously inspired by Pfeuty's relation. 

Let us also take note of two limiting views on the $J_{1}$-$J_{2}$ chain. If $J_{1}$ is strong compared to $J_{2}$, then it is obviously a single chain with the sites $1$ and $L$ as the edges, and $\rho^{x}_{1,L}$ as the end-to-end correlation. It can have only two edge modes. But if $J_{1}$ is quite small compared to $J_{2}$, it can be viewed as a problem of two weakly-coupled chains (see Fig.~\ref{fig:model}). Accordingly, the sites $2$ and $L$ could behave as the edges of the sub-chain of even-numbered sites, while $1$ and $L-1$ would be the ends of the odd sub-chain. In this case, $\rho^{x}_{2,L}=\langle\sigma^{x}_{2}\sigma^{x}_{L}\rangle$ and $\rho^{x}_{1,L-1}=\langle\sigma^{x}_{1}\sigma^{x}_{L-1}\rangle$ may also exhibit $p^{8}$ behavior, which would imply the occurrence of four edge-modes (which is exact for $J_{1}=0$). 

\subsection{FM Phase}
We now calculate the end-to-end and other {\em near-end} spin-spin correlations. Since the exponent of the order parameter $p$ is the object of our study here, we plot the $\log$ of different correlations against $\log{p}$, and compare with the physically motivated $y=mx$ lines, where $x$ is $\log{p}$ and the slope $m$ is 2 for the bulk correlation and 8 for the edge modes.

\begin{figure}[t]
\centering
\includegraphics[width=0.42\textwidth]{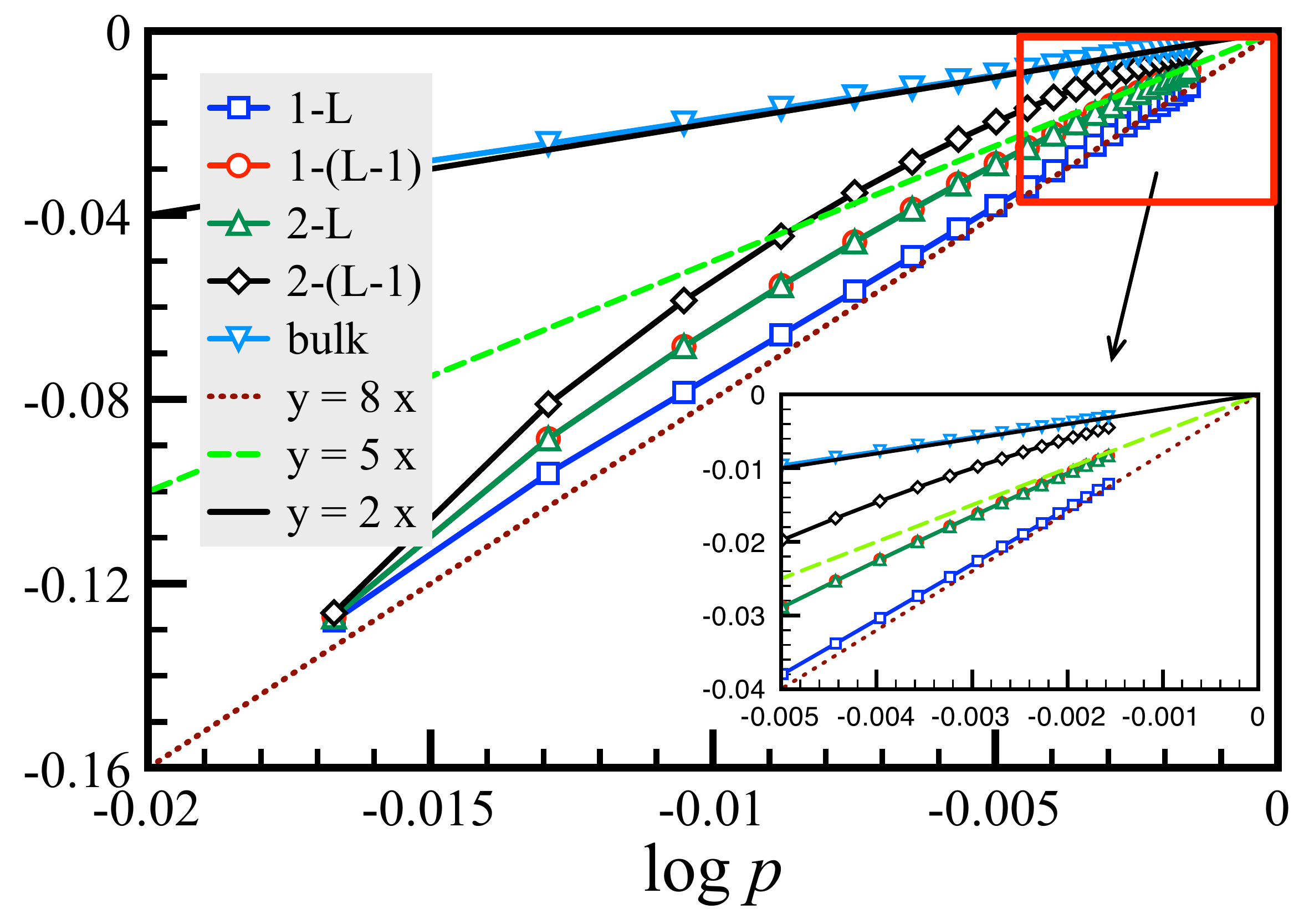} 
\caption{The $\log$ of spin correlations for $J_{2} = -2.0$, generated by varying $J_{1}$, from DMRG calculations for $L=200$. Here, 1-L in the plot legends denotes $\langle \sigma^{x}_{1}\sigma^{x}_{L}\rangle$, and likewise for other near-end correlations. The `bulk' is a correlation between two far away spins that are also far away from the ends.}
\label{fig:corr-dmrg-fm-j2-2p0}
\end{figure}

The data from a DMRG calculation is shown in Fig.~\ref{fig:corr-dmrg-fm-j2-2p0}, where, in addition to $\rho^{x}_{1,L}$, we also plot $\rho^{x}_{1,L-1}$, $\rho^{x}_{2,L}$, $\rho^{x}_{2,L-1}$, and the correlation between two spins deep inside the bulk. This data is parametrically generated by varying $J_{1}$ (from 0 to $-4$) for a fixed $J_{2}$ ($=-2$). While the bulk spin-spin correlation correctly falls on $y=2x$ line, the $\rho^{x}_{1,L}$ data follows $y=8x$. Clearly, the edges behave differently from the bulk. As $\rho^{x}_{1,L}$ follows $p^{8}$ rule, it implies the existence of two edge-modes in the FM phase.

The other notable features in Fig.~\ref{fig:corr-dmrg-fm-j2-2p0} are the behaviours of $\rho^{x}_{1,L-1}$, $\rho^{x}_{2,L}$ and $\rho^{x}_{2,L-1}$. They all start from $y=8x$ line at $J_{1}=0$, and as $J_{1}$ grows more and more negative, they neatly approach $y=2x$ or $5x$ lines. See the inset of Fig.~\ref{fig:corr-dmrg-fm-j2-2p0} for clarity. The spins at sites 2 and $L-1$ are clearly part of the bulk, as $\rho^{x}_{2,L-1}$ tends to $p^{2}$ for strong enough $J_{1}$. Very close to $J_{1} = 0$, the $\rho^{x}_{1,L-1}$ and $\rho^{x}_{2,L}$ behave like $p^{8}$, as expected for the end-to-end spin correlations for the odd and even sub-chains. But when $J_{1}$ grows stronger, they approach $p^{5}$. This suggests that $\langle \sigma^{x}_{1}\rangle=\langle\sigma^{x}_{L}\rangle=p^{4}$, and $\langle \sigma^{x}_{2}\rangle=\langle\sigma^{x}_{L-1}\rangle=p^{4}$ or $p$ if they were to respectively behave as the edges or the bulk. This numerical observation for the local expectations of spins is also consistent with the behaviour of $\rho^{x}_{2,L-1}$ and $\rho^{x}_{1,L}$. 

\begin{figure}[t] 
   \centering
   \includegraphics[width=0.42\textwidth]{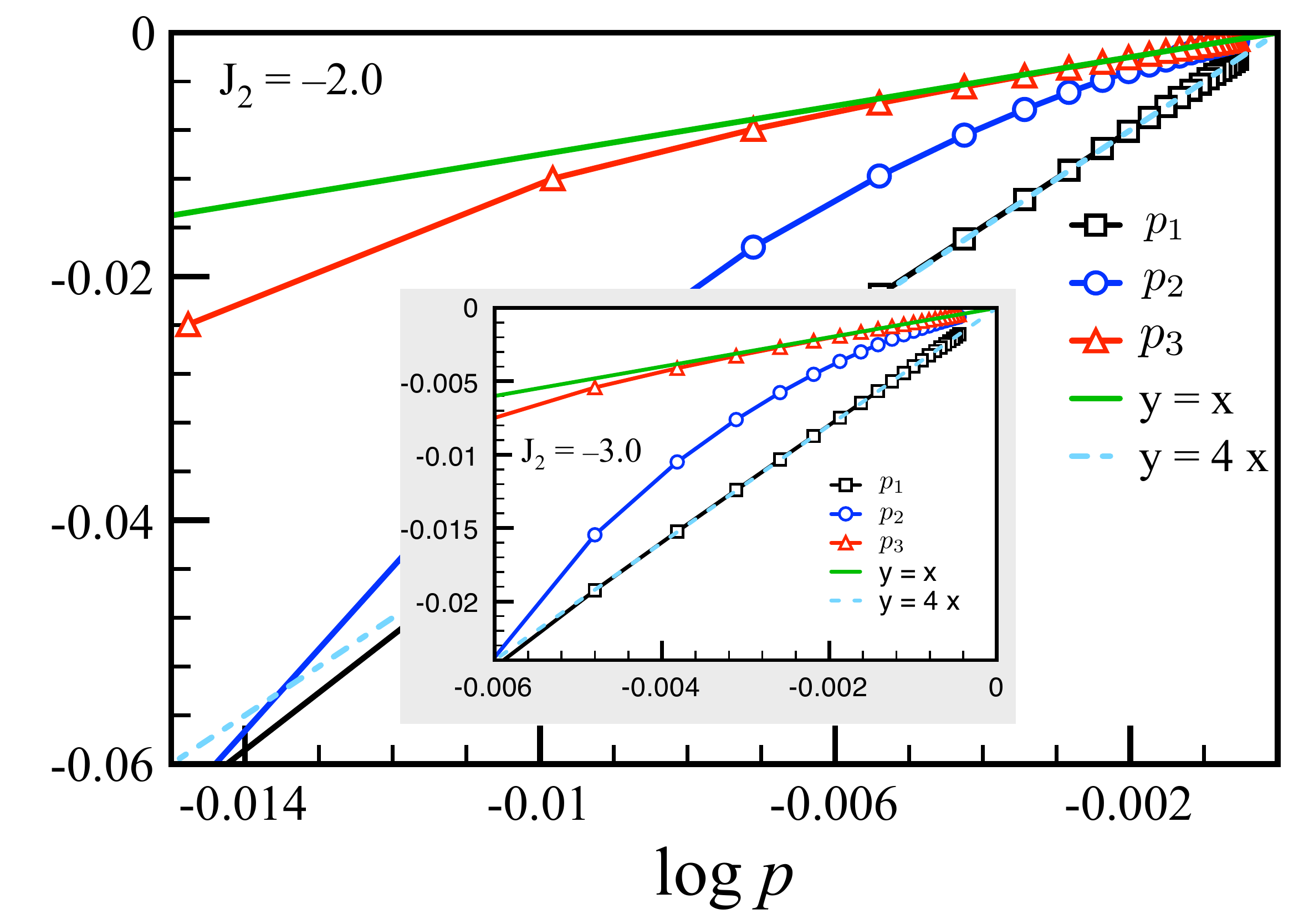}
   \caption{Cluster-MFT combined with DMRG in the FM phase. Here, $p_{i}=\langle \sigma^{x}_{i}\rangle$, and $p$ is the FM order parameter.}
   \label{fig:cmft-fm}
\end{figure}

Motivated by these observations, we also do CMFT calculations. The mean-field model studied for the FM phase is $\Hhat_{FM} = \Hhat +  p[(J_1 + J_2)\sigma^x_L  +  J_2\sigma^x_{L-1}]$, where one end of the cluster is kept free, while the other couples to the `bulk' mean-field, $p$, to be calculated self-consistently. This allows us to look into the local behaviour of the spins near the free-end and compare it with $p$. We do exact numerical diagonalization of $\Hhat_{FM}$ on the chains of lengths up to $16$, and calculate $p_{i}=\langle\sigma^{x}_{i}\rangle$ for the first few spins on the free-end side, that is, $i=1$, 2, 3 etc. In order to access larger chain lengths, we also combine CMFT with DMRG to grow the cluster size. Both approaches give us consistent results. In Fig.~\ref{fig:cmft-fm}, we present one such data. Notably, the $p_{1}$ goes as $p^{4}$, while $p_{2}$ starts from $p^{4}$ (for $J_{1}=0$) and approaches $p$ for strong $J_{1}$ (for different fixed values of $J_{2}$). This is exactly like what we inferred from the DMRG data. Hence, the simple-minded CMFT is consistent with DMRG, and both these calculations provide clear evidence for the occurrence of two edge-modes in the FM phase. The behaviour of $p_{2}$ and $\rho^{x}_{2,L-1}$ also indicates that, very close to $J_{1}=0$, there may occur four edge modes (as in two decoupled sub-chains).

\subsection{DS-AFM Phase}
\begin{figure}[t] 
   \centering
   \includegraphics[width=0.42\textwidth]{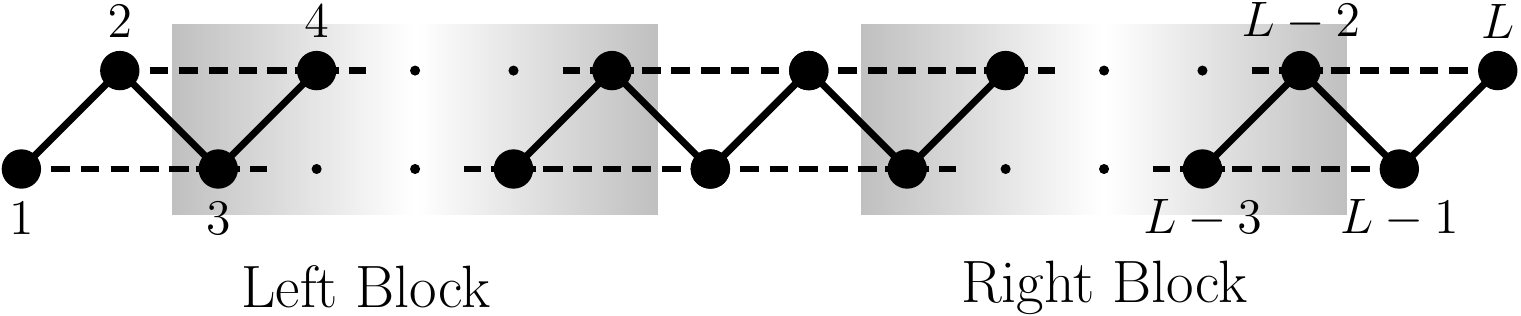} 
   \caption{Free-ends DMRG. While the left and right (shaded) blocks are iteratively renormalised, the two boundary spins (on both sides) are kept un-renormalised.}
   \label{fig:free-end-dmrg}
\end{figure}

We do the same analysis in the DS-AFM phase, except that now we compare different end-to-end spin correlations with different powers of $p^{ }_{ds}$. Since this phase is more frustrated (and fourfold degenerate), we find quite a bit of scatter in the end-to-end correlations in our simple implementation of DMRG (without using parity symmetry). We did not face this difficulty in the FM phase. To improve it, we set up DMRG slightly differently by keeping the spins 1 and 2 at the left-end, and $L-1$ and $L$ at the right-end, free (as in Fig.~\ref{fig:free-end-dmrg}). That is, unlike in the usual DMRG approach wherein all the spins of the left and right blocks are updated, we now keep the last two spins at the left and the right ends un-renormalized, while everything in between is being updated. It is found to give better results. 

\begin{figure}[t]
\centering
\includegraphics[width=0.42\textwidth]{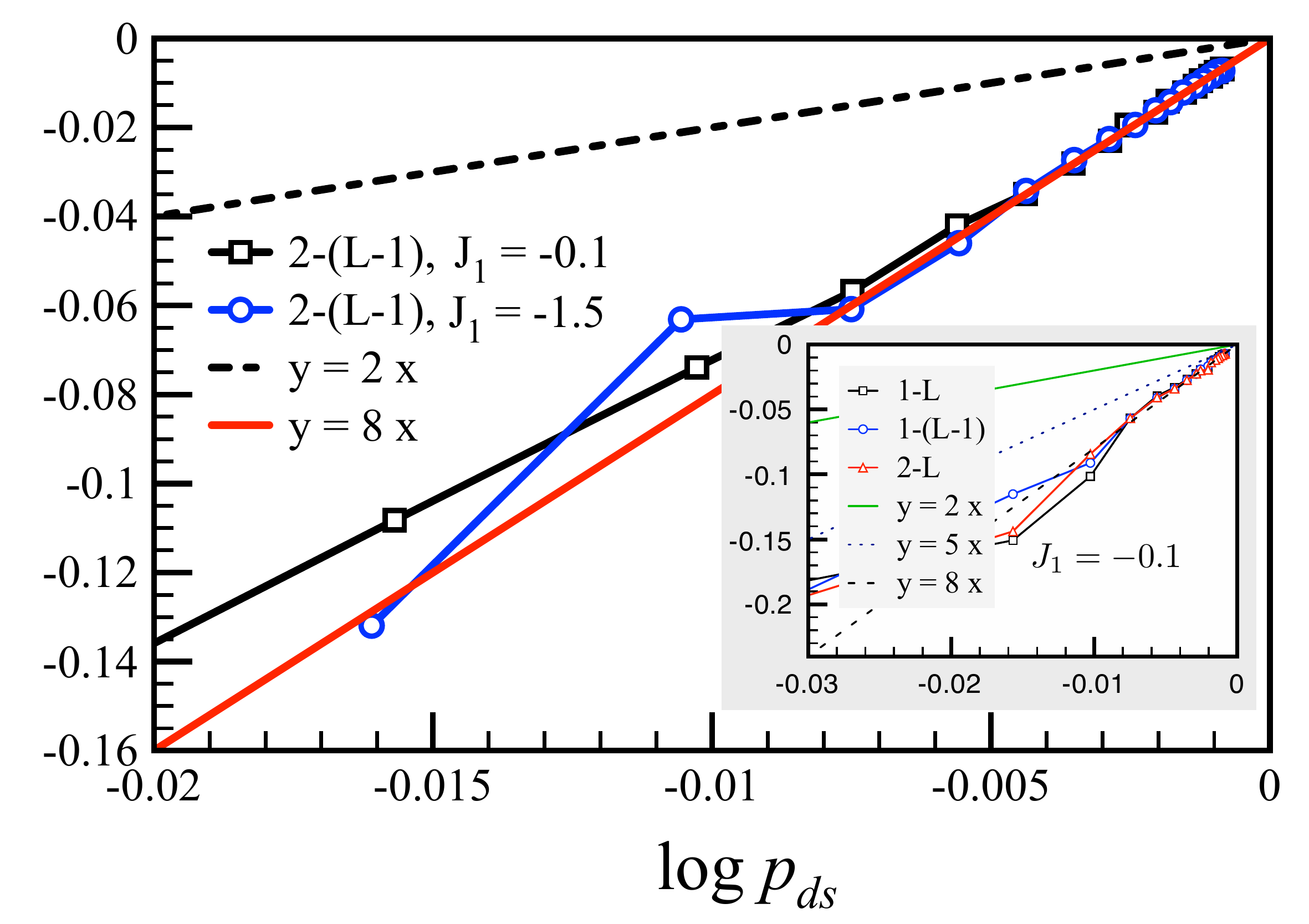} 
\caption{The $\log{\rho^{x}_{2,L-1}}$ from DMRG calculation in the DS-AFM phase for two different $J_{1}$. Here, $J_{2}$ varies from the upper critical line to some larger positive values. (Inset) The $\log$ of $\rho^{x}_{1,L}$, $\rho^{x}_{1,L-1}$ and $\rho^{x}_{2,L}$ vs. $\log{p^{ }_{ds}}$ in the same phase.}
\label{fig:free-end-dmrg-afm}
\end{figure}

The data for different end-to-end correlations from a free-ends DMRG calculation is presented in Fig.~\ref{fig:free-end-dmrg-afm}. This is clearly in accordance with the view that when $J_{2}$ dominates (which is so in the DS-AFM phase), the two sub-chains tend to behave as two. A nice check of this comes from $\rho^{x}_{2,L-1}$. While 1 and $L$ are the actual free-ends of the chain, $2$ and $L-1$ are not. Therefore, if $2$ and $L-1$ were to behave like the free ends (of the respective sub-chains), then $\langle \rho^{x}_{2}\rangle$ and $\langle \rho^{x}_{L-1}\rangle$ should each behave as $p_{ds}^{4}$, or in other words, $\rho^{x}_{2,L-1}$ should behave as $p^{8}_{ds}$. Interestingly, this expectation is precisely met by our numerical data. In Fig.~\ref{fig:free-end-dmrg-afm}, we plot $\log{\rho^{x}_{2,L-1}}$ vs. $\log{p_{ds}}$ for $J_{1}=-0.1$ and $-1.5$. This data is generated by varying $J_{2}$ from the points on the upper critical line to $J_{2}=8$, and it follows $y=8x$ line all along. Moreover, at no point it shows the tendency to go towards $y=2x$ line, that is, of showing the bulk behaviour. We have checked it for different values of $J_{1}$ in the DS-AMF phase. This is unlike the FM phase (see Fig.~\ref{fig:corr-dmrg-fm-j2-2p0}), where $\rho^{x}_{2,L-1}$ can behave as both, depending upon the strength of $J_{1}$ relative to $J_{2}$. Furthermore, $\rho^{x}_{1,L-1}$ and $\rho^{x}_{2,L}$ expectedly approach $p^{8}_{ds}$ behaviour (see the inset). This data suggests that the DM-AFM phase always supports four edge modes.

\begin{figure}[h]
   \centering
   \includegraphics[width=0.42\textwidth]{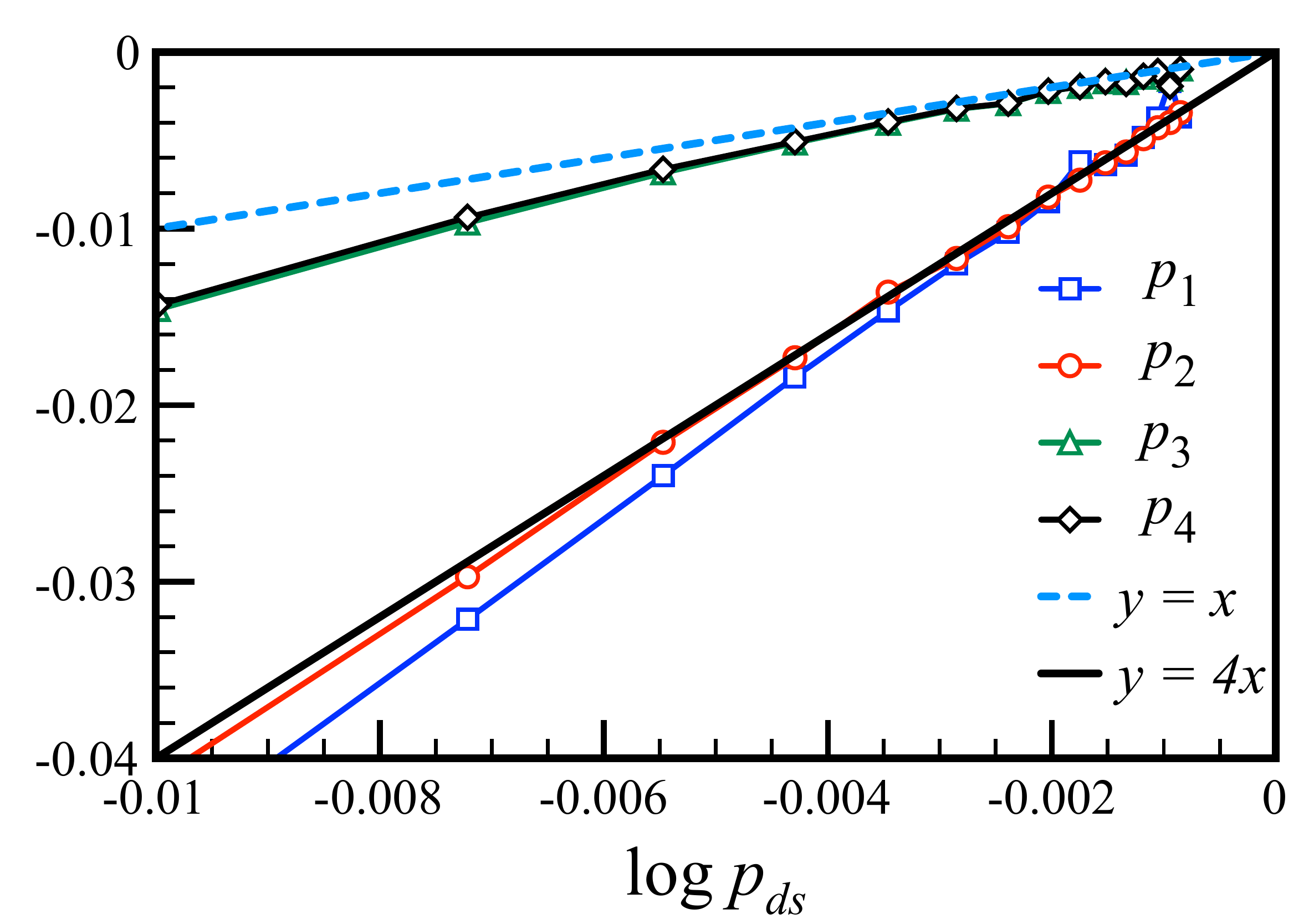}
   \caption{$\log{p_{i}}$ vs. $\log{p_{ds}}$ from CMFT calculations ($L = 200$).}
   \label{fig:cmft-dmrg-afm}
\end{figure}

\begin{figure}[t] 
   \centering
   \includegraphics[width=0.42\textwidth]{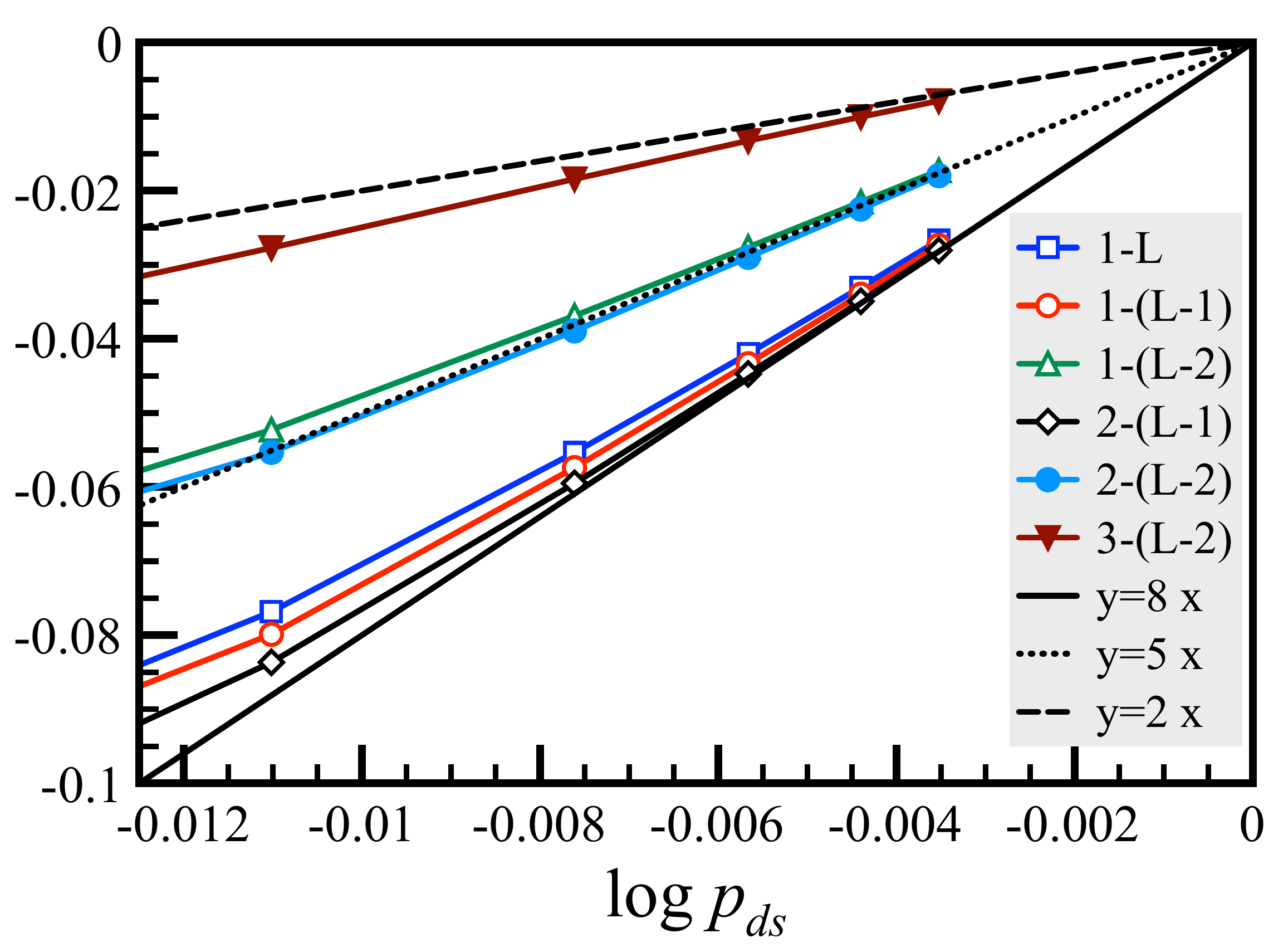} 
   \caption{Exact numerical diagonalization results for $\log{p_{i}}$ in the ground state of $\Hhat$ for 16 spins. Here, $J_{1} = -0.1$ and data is collected by changing $J_{2}$ in the positive direction in the DS-AFM phase. }
   \label{fig:ed16} 
\end{figure}

In Fig.~\ref{fig:cmft-dmrg-afm}, we present the data for the local spin expectation values, $\{p_{i}$\}, near the free end from a CMFT calculation (combined with DMRG). The mean-field model in this case is $\Hhat_{DS-AFM} = \Hhat + p_{ds}[\pm (J_1 + J_2)\sigma^x_L  \pm  J_2\sigma^x_{L-1}]$, where four possible combinations of the signs correspond to four double-staggered mean-field orders, and they all give the same result. The order parameter $p_{ds}$ is determined self-consistently. While $p_{3}$ and $p_{4}$ approach $p_{ds}$ (as expected for the spins in the bulk), the spin expectations at two sub-chain edges, $p_{1}$ and $p_{2}$, exhibit $p_{ds}^{4}$ behaviour, consistent with the DMRG data in Fig.~\ref{fig:free-end-dmrg-afm}. We also do an exact numerical diagonalization (without mean-field) calculation of $\Hhat$ with 16 spins. Technically $p_{ds}$ is zero here. But we take the square root of the modulus of the correlation two spins away from the edges and from each other as $p_{ds}$, and compare it with different end-to-end correlations (see Fig.~\ref{fig:ed16}). Even this simple calculation reveals a behaviour that is consistent with DMRG and CMFT. All of these calculations clearly ascertain that the frustrated DS-AFM phase of the $J_{1}$-$J_{2}$ QI chain indeed supports four edge modes.

\section{\label{sec:sum} Summary}
\begin{figure}[t] 
   \centering
   \includegraphics[width=0.42\textwidth]{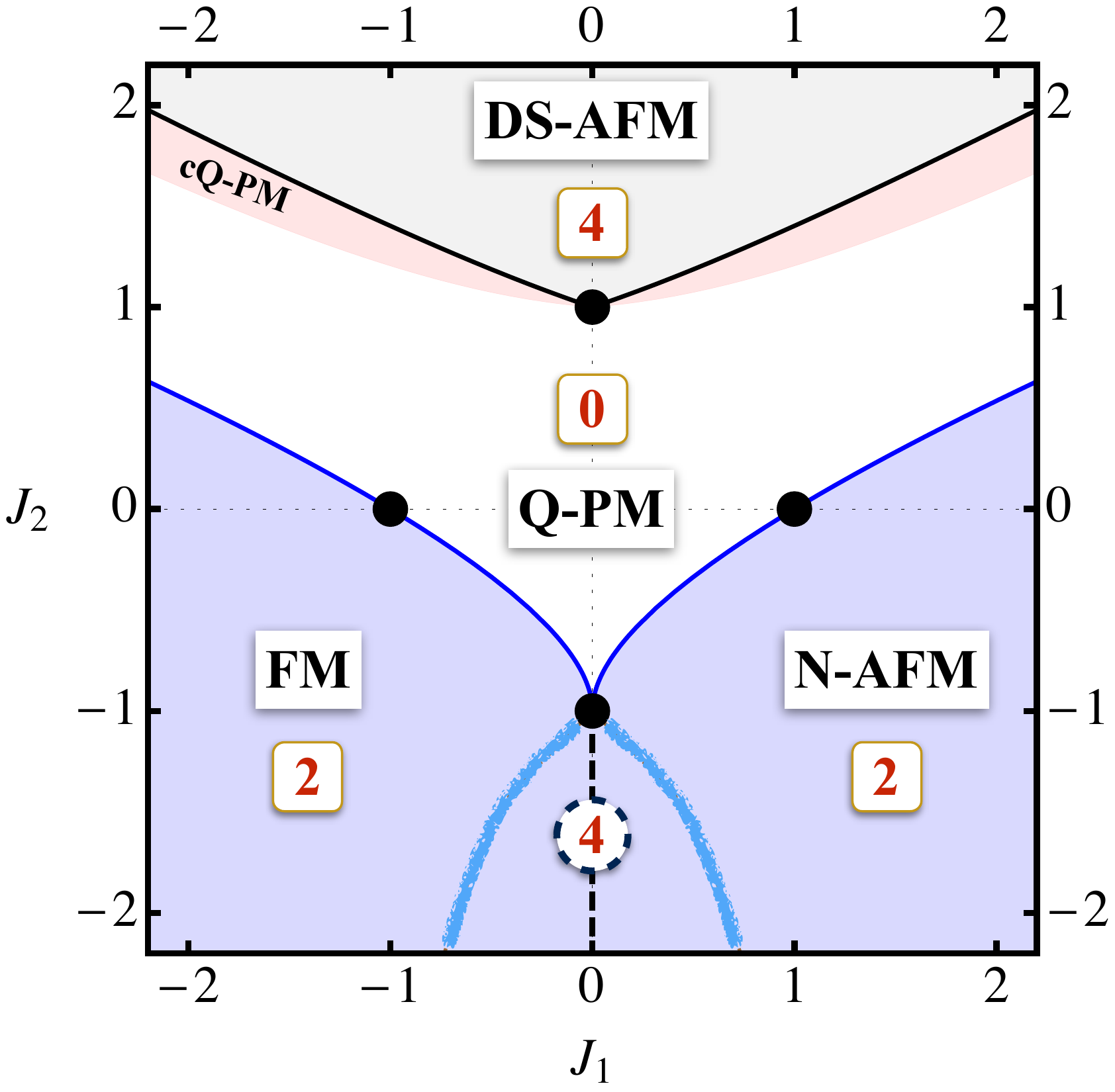} 
   \caption{The quantum phase diagram of the $J_{1}$-$J_{2}$ Ising chain in a transverse field ($h=1$), characterised in terms of the number of edge modes in each phase.}
   \label{fig:summary}
\end{figure}

Our basic motivation behind this work was to study the effect of competing Ising interactions on the occurrence of the edge-modes in a quantum Ising chain. This we have done by numerically investigating the ground state properties of the one-dimensional $J_{1}$-$J_{2}$ Ising model in a transverse field, that is, the $\Hhat$ of Eq.~(\ref{eq:model}). By doing systematic DMRG and CMFT calculations on $\Hhat$, we have generated its quantum phase diagram, and ascertained the presence of the edge modes in the ordered phases. The outcome of this effort is summarised in Fig.~\ref{fig:summary}.

There are two ordered regions in the ground state of $\Hhat$, one of which shows twofold degenerate ferromagnetic (FM) or N\'eel antiferromagnetic order (N-AFM) for negative or positive values of $J_{1}$, respectively.  The other phase, dominated by $J_{2}$, exhibits fourfold degenerate double-staggered antiferromagnetic (DS-AFM) order. In between lies a quantum disordered region consisting mostly of a gapped paramagnetic phase (Q-PM), and a small gapless critical phase (cQ-PM) adjacent to the DS-AFM phase. There are four exact quantum critical points, $(J_{1},J_{2})=(\pm 1,0)$ and $(0,\pm 1)$, shown as filled black circles in Fig.~\ref{fig:summary}. The dashed, $J_{1}=0$ and $J_{2}<-1$, black line is the exact level-crossing line between the FM and N-AFM phases. The rest of the phase boundaries are generated numerically. These phase boundaries in Fig.~\ref{fig:summary} have been smoothened by empirically fitting the data of Fig.~\ref{fig:QPD} to $\sqrt{1+a|x|+bx^{2}}$ for the DS-AFM and cQ-PM phase boundaries, and $-1+ c|x|+(1-c)\sqrt{|x|}$ for the FM phase boundary. These forms are guided by the data itself, and are constrained to pass through four exact critical points, and to become linear for large $x$. One can further improve it, but here we keep it simple.

We characterise these ordered phases in terms of the edges modes that occur therein. We compute different end-to-end spin-spin correlations on the chains with open boundaries, and compare them with the $8^{th}$, $5^{th}$ and $2^{nd}$ power of the order parameter. Sufficiently inside the ordered phase, if an end-to-end correlation goes as the $8^{th}$ power of the order parameter, then it indicates the occurrence of two edge modes. If it goes as the $2^{nd}$ power, then it is clearly a bulk behaviour, and implies no edge modes. The $5^{th}$ power behaviour indicates that one of the two concerned sites supports a Majorana-like edge mode, while the other behaves as bulk. These empirical rules are inspired by the exact $8^{th}$ power behaviour for the end-to-end  correlation in the nearest-neighbor QI chain, and are born out well by our numerical calculations. 

Thus, we come to conclude that the DS-AFM phase supports four edge modes (indicated by a 4 inside the little box in Fig.~\ref{fig:summary}). There are no edge-modes in the quantum paramagnetic phases (Q-PM as well as cQ-PM). The FM and N-AFM phases support two edge modes for most parts. However, very near the dashed $J_{1}=0$ line that has four edge modes (indicated by a 4 inside the circle), our calculations seem to suggest that the FM and N-AFM phases may also realise four edge modes. This possible cross-over from two to four edge modes is denoted by two thick hazy light-blue lines surrounding the $J_{1}=0$ dashed line, and is roughly estimated by $\log{\rho^{x}_{2,L-1}}/\log{p}$ going above $7$ as $J_{2}$ grows more and more negative for a small $J_{1}$. Notably, the number of edge modes is also same as the degeneracy of the ordered phase. It roughly makes sense, because a pair of Majorana modes leads to twofold degeneracy, therefore, two pairs of these edge modes would cause fourfold degeneracy. In view of this, a more careful analysis needs to be done to check whether the cross-over from two to four edge modes in the FM/N-AFM phase indeed occurs.

These are indirect inferences on the edge modes. A more direct study in the fermionized form ($\Hhat_{1}$) is most desirable. We did attempt a fermionic mean-field calculation that looked okay well inside the FM/N-AFM phase, because the decoupling of the $J_{2}$ interaction generates $J_{1}$ like terms. But it wasn't satisfactory for the DS-AFM phase, and close to $J_{1}=0$. We are looking for alternative ways to study the corresponding fermion model to see the Majorana edge modes in the $J_{1}$-$J_{2}$ QI chain.

\begin{acknowledgments}
SJ acknowledges CSIR for financial support. We also acknowledge DST-FIST support for the computing facility in SPS.
\end{acknowledgments}
\bibliography{references}

\begin{thebibliography}{28}%
\makeatletter
\providecommand \@ifxundefined [1]{%
 \@ifx{#1\undefined}
}%
\providecommand \@ifnum [1]{%
 \ifnum #1\expandafter \@firstoftwo
 \else \expandafter \@secondoftwo
 \fi
}%
\providecommand \@ifx [1]{%
 \ifx #1\expandafter \@firstoftwo
 \else \expandafter \@secondoftwo
 \fi
}%
\providecommand \natexlab [1]{#1}%
\providecommand \enquote  [1]{``#1''}%
\providecommand \bibnamefont  [1]{#1}%
\providecommand \bibfnamefont [1]{#1}%
\providecommand \citenamefont [1]{#1}%
\providecommand \href@noop [0]{\@secondoftwo}%
\providecommand \href [0]{\begingroup \@sanitize@url \@href}%
\providecommand \@href[1]{\@@startlink{#1}\@@href}%
\providecommand \@@href[1]{\endgroup#1\@@endlink}%
\providecommand \@sanitize@url [0]{\catcode `\\12\catcode `\$12\catcode
  `\&12\catcode `\#12\catcode `\^12\catcode `\_12\catcode `\%12\relax}%
\providecommand \@@startlink[1]{}%
\providecommand \@@endlink[0]{}%
\providecommand \url  [0]{\begingroup\@sanitize@url \@url }%
\providecommand \@url [1]{\endgroup\@href {#1}{\urlprefix }}%
\providecommand \urlprefix  [0]{URL }%
\providecommand \Eprint [0]{\href }%
\providecommand \doibase [0]{http://dx.doi.org/}%
\providecommand \selectlanguage [0]{\@gobble}%
\providecommand \bibinfo  [0]{\@secondoftwo}%
\providecommand \bibfield  [0]{\@secondoftwo}%
\providecommand \translation [1]{[#1]}%
\providecommand \BibitemOpen [0]{}%
\providecommand \bibitemStop [0]{}%
\providecommand \bibitemNoStop [0]{.\EOS\space}%
\providecommand \EOS [0]{\spacefactor3000\relax}%
\providecommand \BibitemShut  [1]{\csname bibitem#1\endcsname}%
\let\auto@bib@innerbib\@empty
\bibitem [{\citenamefont {de~Gennes}(1963)}]{deGennes}%
  \BibitemOpen
  \bibfield  {author} {\bibinfo {author} {\bibfnamefont {P.~G.}\ \bibnamefont
  {de~Gennes}},\ }\href@noop {} {\bibfield  {journal} {\bibinfo  {journal}
  {Solid State Commun}\ }\textbf {\bibinfo {volume} {1}},\ \bibinfo {pages}
  {132} (\bibinfo {year} {1963})}\BibitemShut {NoStop}%
\bibitem [{\citenamefont {Chakrabarti}\ \emph {et~al.}(1996)\citenamefont
  {Chakrabarti}, \citenamefont {Dutta},\ and\ \citenamefont
  {Sen}}]{book.Bikas}%
  \BibitemOpen
  \bibfield  {author} {\bibinfo {author} {\bibfnamefont {B.~K.}\ \bibnamefont
  {Chakrabarti}}, \bibinfo {author} {\bibfnamefont {A.}~\bibnamefont {Dutta}},
  \ and\ \bibinfo {author} {\bibfnamefont {P.}~\bibnamefont {Sen}},\
  }\href@noop {} {\emph {\bibinfo {title} {Quantum Ising Phases and Transitions
  in Transverse Ising models}}}\ (\bibinfo  {publisher} {Springer-Verlag,
  Berlin},\ \bibinfo {year} {1996})\BibitemShut {NoStop}%
\bibitem [{\citenamefont {Fisher}(1995)}]{randomQI.Fisher}%
  \BibitemOpen
  \bibfield  {author} {\bibinfo {author} {\bibfnamefont {D.~S.}\ \bibnamefont
  {Fisher}},\ }\href@noop {} {\bibfield  {journal} {\bibinfo  {journal} {Phys.
  Rev. B}\ }\textbf {\bibinfo {volume} {51}},\ \bibinfo {pages} {6411}
  (\bibinfo {year} {1995})}\BibitemShut {NoStop}%
\bibitem [{\citenamefont {Moessner}\ \emph {et~al.}(2000)\citenamefont
  {Moessner}, \citenamefont {Sondhi},\ and\ \citenamefont
  {Chandra}}]{Moessner}%
  \BibitemOpen
  \bibfield  {author} {\bibinfo {author} {\bibfnamefont {R.}~\bibnamefont
  {Moessner}}, \bibinfo {author} {\bibfnamefont {S.~L.}\ \bibnamefont
  {Sondhi}}, \ and\ \bibinfo {author} {\bibfnamefont {P.}~\bibnamefont
  {Chandra}},\ }\href@noop {} {\bibfield  {journal} {\bibinfo  {journal} {Phys.
  Rev. Lett.}\ }\textbf {\bibinfo {volume} {84}},\ \bibinfo {pages} {4457}
  (\bibinfo {year} {2000})}\BibitemShut {NoStop}%
\bibitem [{\citenamefont {Chakraborty}\ \emph {et~al.}(2004)\citenamefont
  {Chakraborty}, \citenamefont {Henelius}, \citenamefont {Kj{\o}nsberg},
  \citenamefont {Sandvik},\ and\ \citenamefont {Girvin}}]{QI.Holo.Girvin}%
  \BibitemOpen
  \bibfield  {author} {\bibinfo {author} {\bibfnamefont {P.~B.}\ \bibnamefont
  {Chakraborty}}, \bibinfo {author} {\bibfnamefont {P.}~\bibnamefont
  {Henelius}}, \bibinfo {author} {\bibfnamefont {H.}~\bibnamefont
  {Kj{\o}nsberg}}, \bibinfo {author} {\bibfnamefont {A.~W.}\ \bibnamefont
  {Sandvik}}, \ and\ \bibinfo {author} {\bibfnamefont {S.~M.}\ \bibnamefont
  {Girvin}},\ }\href@noop {} {\bibfield  {journal} {\bibinfo  {journal} {Phys.
  Rev. B}\ }\textbf {\bibinfo {volume} {70}},\ \bibinfo {pages} {144411}
  (\bibinfo {year} {2004})}\BibitemShut {NoStop}%
\bibitem [{\citenamefont {Coldea}\ \emph {et~al.}(2010)\citenamefont {Coldea},
  \citenamefont {Tennant}, \citenamefont {Wheeler}, \citenamefont {Wawrzynska},
  \citenamefont {Prabhakaran}, \citenamefont {Telling}, \citenamefont
  {Habicht}, \citenamefont {Smeibidl},\ and\ \citenamefont {Kiefer}}]{Coldea}%
  \BibitemOpen
  \bibfield  {author} {\bibinfo {author} {\bibfnamefont {R.}~\bibnamefont
  {Coldea}}, \bibinfo {author} {\bibfnamefont {D.~A.}\ \bibnamefont {Tennant}},
  \bibinfo {author} {\bibfnamefont {E.~M.}\ \bibnamefont {Wheeler}}, \bibinfo
  {author} {\bibfnamefont {E.}~\bibnamefont {Wawrzynska}}, \bibinfo {author}
  {\bibfnamefont {D.}~\bibnamefont {Prabhakaran}}, \bibinfo {author}
  {\bibfnamefont {M.}~\bibnamefont {Telling}}, \bibinfo {author} {\bibfnamefont
  {K.}~\bibnamefont {Habicht}}, \bibinfo {author} {\bibfnamefont
  {P.}~\bibnamefont {Smeibidl}}, \ and\ \bibinfo {author} {\bibfnamefont
  {K.}~\bibnamefont {Kiefer}},\ }\href@noop {} {\bibfield  {journal} {\bibinfo
  {journal} {Science}\ }\textbf {\bibinfo {volume} {327}},\ \bibinfo {pages}
  {177} (\bibinfo {year} {2010})}\BibitemShut {NoStop}%
\bibitem [{\citenamefont {Bermudez}\ \emph {et~al.}(2012)\citenamefont
  {Bermudez}, \citenamefont {Almeida}, \citenamefont {Ott}, \citenamefont
  {Kaufmann}, \citenamefont {Ulm}, \citenamefont {Poschinger}, \citenamefont
  {Schmidt-Kaler}, \citenamefont {Retzker},\ and\ \citenamefont
  {Plenio}}]{bermudez-qi-expt}%
  \BibitemOpen
  \bibfield  {author} {\bibinfo {author} {\bibfnamefont {A.}~\bibnamefont
  {Bermudez}}, \bibinfo {author} {\bibfnamefont {J.}~\bibnamefont {Almeida}},
  \bibinfo {author} {\bibfnamefont {K.}~\bibnamefont {Ott}}, \bibinfo {author}
  {\bibfnamefont {H.}~\bibnamefont {Kaufmann}}, \bibinfo {author}
  {\bibfnamefont {S.}~\bibnamefont {Ulm}}, \bibinfo {author} {\bibfnamefont
  {U.}~\bibnamefont {Poschinger}}, \bibinfo {author} {\bibfnamefont
  {F.}~\bibnamefont {Schmidt-Kaler}}, \bibinfo {author} {\bibfnamefont
  {A.}~\bibnamefont {Retzker}}, \ and\ \bibinfo {author} {\bibfnamefont
  {M.~B.}\ \bibnamefont {Plenio}},\ }\href
  {http://stacks.iop.org/1367-2630/14/i=9/a=093042} {\bibfield  {journal}
  {\bibinfo  {journal} {New J. Phys.}\ }\textbf {\bibinfo {volume} {14}},\
  \bibinfo {pages} {093042} (\bibinfo {year} {2012})}\BibitemShut {NoStop}%
\bibitem [{\citenamefont {Wenzel}\ \emph {et~al.}(2012)\citenamefont {Wenzel},
  \citenamefont {Coletta}, \citenamefont {Korshunov},\ and\ \citenamefont
  {Mila}}]{Mila.Square}%
  \BibitemOpen
  \bibfield  {author} {\bibinfo {author} {\bibfnamefont {S.}~\bibnamefont
  {Wenzel}}, \bibinfo {author} {\bibfnamefont {T.}~\bibnamefont {Coletta}},
  \bibinfo {author} {\bibfnamefont {S.~E.}\ \bibnamefont {Korshunov}}, \ and\
  \bibinfo {author} {\bibfnamefont {F.}~\bibnamefont {Mila}},\ }\href {\doibase
  10.1103/PhysRevLett.109.187202} {\bibfield  {journal} {\bibinfo  {journal}
  {Phys. Rev. Lett.}\ }\textbf {\bibinfo {volume} {109}},\ \bibinfo {pages}
  {187202} (\bibinfo {year} {2012})}\BibitemShut {NoStop}%
\bibitem [{\citenamefont {Kumar}\ and\ \citenamefont
  {Jalal}(2013)}]{bkumar.somenath}%
  \BibitemOpen
  \bibfield  {author} {\bibinfo {author} {\bibfnamefont {B.}~\bibnamefont
  {Kumar}}\ and\ \bibinfo {author} {\bibfnamefont {S.}~\bibnamefont {Jalal}},\
  }\href {\doibase 10.1103/PhysRevA.88.011802} {\bibfield  {journal} {\bibinfo
  {journal} {Phys. Rev. A}\ }\textbf {\bibinfo {volume} {88}},\ \bibinfo
  {pages} {011802(R)} (\bibinfo {year} {2013})}\BibitemShut {NoStop}%
\bibitem [{\citenamefont {Pfeuty}(1970)}]{pfeuty}%
  \BibitemOpen
  \bibfield  {author} {\bibinfo {author} {\bibfnamefont {P.}~\bibnamefont
  {Pfeuty}},\ }\href@noop {} {\bibfield  {journal} {\bibinfo  {journal} {Annals
  of Physics}\ }\textbf {\bibinfo {volume} {57}},\ \bibinfo {pages} {79}
  (\bibinfo {year} {1970})}\BibitemShut {NoStop}%
\bibitem [{\citenamefont {Kitaev}(2001)}]{Kitaev.QWire}%
  \BibitemOpen
  \bibfield  {author} {\bibinfo {author} {\bibfnamefont {A.~Y.}\ \bibnamefont
  {Kitaev}},\ }\href {http://stacks.iop.org/1063-7869/44/i=10S/a=S29}
  {\bibfield  {journal} {\bibinfo  {journal} {Physics-Uspekhi}\ }\textbf
  {\bibinfo {volume} {44}},\ \bibinfo {pages} {131} (\bibinfo {year}
  {2001})}\BibitemShut {NoStop}%
\bibitem [{\citenamefont {Nayak}\ \emph {et~al.}(2008)\citenamefont {Nayak},
  \citenamefont {Simon}, \citenamefont {Stern}, \citenamefont {Freedman},\ and\
  \citenamefont {Das~Sarma}}]{Nayak.RMP}%
  \BibitemOpen
  \bibfield  {author} {\bibinfo {author} {\bibfnamefont {C.}~\bibnamefont
  {Nayak}}, \bibinfo {author} {\bibfnamefont {S.~H.}\ \bibnamefont {Simon}},
  \bibinfo {author} {\bibfnamefont {A.}~\bibnamefont {Stern}}, \bibinfo
  {author} {\bibfnamefont {M.}~\bibnamefont {Freedman}}, \ and\ \bibinfo
  {author} {\bibfnamefont {S.}~\bibnamefont {Das~Sarma}},\ }\href {\doibase
  10.1103/RevModPhys.80.1083} {\bibfield  {journal} {\bibinfo  {journal} {Rev.
  Mod. Phys.}\ }\textbf {\bibinfo {volume} {80}},\ \bibinfo {pages} {1083}
  (\bibinfo {year} {2008})}\BibitemShut {NoStop}%
\bibitem [{\citenamefont {Alicea}(2012)}]{jason-majorana}%
  \BibitemOpen
  \bibfield  {author} {\bibinfo {author} {\bibfnamefont {J.}~\bibnamefont
  {Alicea}},\ }\href {\doibase 10.1088/0034-4885/75/7/076501} {\bibfield
  {journal} {\bibinfo  {journal} {Rep. Prog. Phys}\ }\textbf {\bibinfo {volume}
  {75}},\ \bibinfo {pages} {076501} (\bibinfo {year} {2012})}\BibitemShut
  {NoStop}%
\bibitem [{\citenamefont {Barber}\ and\ \citenamefont
  {Duxbury}(1981)}]{QIj1j2.Barber}%
  \BibitemOpen
  \bibfield  {author} {\bibinfo {author} {\bibfnamefont {M.~N.}\ \bibnamefont
  {Barber}}\ and\ \bibinfo {author} {\bibfnamefont {P.~M.}\ \bibnamefont
  {Duxbury}},\ }\href@noop {} {\bibfield  {journal} {\bibinfo  {journal} {J.
  Phys. A: Math. Gen.}\ }\textbf {\bibinfo {volume} {14}},\ \bibinfo {pages}
  {L251} (\bibinfo {year} {1981})}\BibitemShut {NoStop}%
\bibitem [{\citenamefont {Selke}(1988)}]{ANNNI.Selke}%
  \BibitemOpen
  \bibfield  {author} {\bibinfo {author} {\bibfnamefont {W.}~\bibnamefont
  {Selke}},\ }\href@noop {} {\bibfield  {journal} {\bibinfo  {journal} {Phys.
  Reports}\ }\textbf {\bibinfo {volume} {170}},\ \bibinfo {pages} {213}
  (\bibinfo {year} {1988})}\BibitemShut {NoStop}%
\bibitem [{Note1()}]{Note1}%
  \BibitemOpen
  \bibinfo {note} {The Majumdar-Ghosh model is a historic 1D problem of
  frustrated quantum (Heisenberg) spin-1/2's~\cite {MG}. Since it precisely
  corresponds to $J_2=J_1/2$, it has become common to refer to this relation
  between $J_1$ and $J_2$ as MG point or line (depending upon the
  context).}\BibitemShut {Stop}%
\bibitem [{Note2()}]{Note2}%
  \BibitemOpen
  \bibinfo {note} {The double-staggered AFM here is same as the `antiphase' in
  the conventional ANNNI literature.}\BibitemShut {Stop}%
\bibitem [{\citenamefont {Sen}\ and\ \citenamefont
  {Chakrabarti}(1991)}]{QIj1j2.Bikas}%
  \BibitemOpen
  \bibfield  {author} {\bibinfo {author} {\bibfnamefont {P.}~\bibnamefont
  {Sen}}\ and\ \bibinfo {author} {\bibfnamefont {B.~K.}\ \bibnamefont
  {Chakrabarti}},\ }\href@noop {} {\bibfield  {journal} {\bibinfo  {journal}
  {Phys. Rev. B}\ }\textbf {\bibinfo {volume} {43}},\ \bibinfo {pages} {13559}
  (\bibinfo {year} {1991})}\BibitemShut {NoStop}%
\bibitem [{\citenamefont {Guimar$\tilde{\rm a}$es}\ \emph
  {et~al.}(2002)\citenamefont {Guimar$\tilde{\rm a}$es}, \citenamefont
  {Plascak}, \citenamefont {SaBarreto},\ and\ \citenamefont
  {Florencio}}]{QIj1j2.Brazil}%
  \BibitemOpen
  \bibfield  {author} {\bibinfo {author} {\bibfnamefont {P.~R.~C.}\
  \bibnamefont {Guimar$\tilde{\rm a}$es}}, \bibinfo {author} {\bibfnamefont
  {J.}~\bibnamefont {Plascak}}, \bibinfo {author} {\bibfnamefont {F.~C.}\
  \bibnamefont {SaBarreto}}, \ and\ \bibinfo {author} {\bibfnamefont
  {J.}~\bibnamefont {Florencio}},\ }\href@noop {} {\bibfield  {journal}
  {\bibinfo  {journal} {Phys. Rev. B}\ }\textbf {\bibinfo {volume} {66}},\
  \bibinfo {pages} {064413} (\bibinfo {year} {2002})}\BibitemShut {NoStop}%
\bibitem [{\citenamefont {Chandra}\ and\ \citenamefont
  {Dasgupta}(2007)}]{QIj1j2.Subinay}%
  \BibitemOpen
  \bibfield  {author} {\bibinfo {author} {\bibfnamefont {A.~K.}\ \bibnamefont
  {Chandra}}\ and\ \bibinfo {author} {\bibfnamefont {S.}~\bibnamefont
  {Dasgupta}},\ }\href@noop {} {\bibfield  {journal} {\bibinfo  {journal}
  {Phys. Rev. E}\ }\textbf {\bibinfo {volume} {75}},\ \bibinfo {pages} {021105}
  (\bibinfo {year} {2007})}\BibitemShut {NoStop}%
\bibitem [{\citenamefont {Beccaria}\ \emph {et~al.}(2007)\citenamefont
  {Beccaria}, \citenamefont {Campostrini},\ and\ \citenamefont
  {Feo}}]{QIj1j2.DMRG}%
  \BibitemOpen
  \bibfield  {author} {\bibinfo {author} {\bibfnamefont {M.}~\bibnamefont
  {Beccaria}}, \bibinfo {author} {\bibfnamefont {M.}~\bibnamefont
  {Campostrini}}, \ and\ \bibinfo {author} {\bibfnamefont {A.}~\bibnamefont
  {Feo}},\ }\href@noop {} {\bibfield  {journal} {\bibinfo  {journal} {Phys.
  Rev. B}\ }\textbf {\bibinfo {volume} {76}},\ \bibinfo {pages} {094410}
  (\bibinfo {year} {2007})}\BibitemShut {NoStop}%
\bibitem [{\citenamefont {Nagy}(2011)}]{QIj1j2.Adam}%
  \BibitemOpen
  \bibfield  {author} {\bibinfo {author} {\bibfnamefont {A.}~\bibnamefont
  {Nagy}},\ }\href@noop {} {\bibfield  {journal} {\bibinfo  {journal} {New J.
  Phys.}\ }\textbf {\bibinfo {volume} {13}},\ \bibinfo {pages} {023015}
  (\bibinfo {year} {2011})}\BibitemShut {NoStop}%
\bibitem [{\citenamefont {Dutta}\ \emph {et~al.}(2010)\citenamefont {Dutta},
  \citenamefont {Divakaran}, \citenamefont {Sen}, \citenamefont {Chakrabarti},
  \citenamefont {Rosenbaum},\ and\ \citenamefont {Aeppli}}]{review.dutta}%
  \BibitemOpen
  \bibfield  {author} {\bibinfo {author} {\bibfnamefont {A.}~\bibnamefont
  {Dutta}}, \bibinfo {author} {\bibfnamefont {U.}~\bibnamefont {Divakaran}},
  \bibinfo {author} {\bibfnamefont {D.}~\bibnamefont {Sen}}, \bibinfo {author}
  {\bibfnamefont {B.~K.}\ \bibnamefont {Chakrabarti}}, \bibinfo {author}
  {\bibfnamefont {T.~F.}\ \bibnamefont {Rosenbaum}}, \ and\ \bibinfo {author}
  {\bibfnamefont {G.}~\bibnamefont {Aeppli}},\ }\href@noop {} {} (\bibinfo
  {year} {2010}),\ \bibinfo {note} {http://arxiv.org/abs/1012.0653}\BibitemShut
  {NoStop}%
\bibitem [{\citenamefont {White}(1993)}]{dmrg1}%
  \BibitemOpen
  \bibfield  {author} {\bibinfo {author} {\bibfnamefont {S.~R.}\ \bibnamefont
  {White}},\ }\href {\doibase 10.1103/PhysRevB.48.10345} {\bibfield  {journal}
  {\bibinfo  {journal} {Phys. Rev. B}\ }\textbf {\bibinfo {volume} {48}},\
  \bibinfo {pages} {10345} (\bibinfo {year} {1993})}\BibitemShut {NoStop}%
\bibitem [{\citenamefont {Schollw\"ock}(2005)}]{dmrg2}%
  \BibitemOpen
  \bibfield  {author} {\bibinfo {author} {\bibfnamefont {U.}~\bibnamefont
  {Schollw\"ock}},\ }\href@noop {} {\bibfield  {journal} {\bibinfo  {journal}
  {Rev. Mod. Phys.}\ }\textbf {\bibinfo {volume} {77}},\ \bibinfo {pages} {259}
  (\bibinfo {year} {2005})}\BibitemShut {NoStop}%
\bibitem [{Note3()}]{Note3}%
  \BibitemOpen
  \bibinfo {note} {In the ANNNI literature, this phase is called the `floating'
  phase, due to the associated physical picture in the classical
  context}\BibitemShut {NoStop}%
\bibitem [{\citenamefont {Niu}\ \emph {et~al.}(2012)\citenamefont {Niu},
  \citenamefont {Chung}, \citenamefont {Hsu}, \citenamefont {Mandal},
  \citenamefont {Raghu},\ and\ \citenamefont {Chakravarty}}]{niu2012prb}%
  \BibitemOpen
  \bibfield  {author} {\bibinfo {author} {\bibfnamefont {Y.}~\bibnamefont
  {Niu}}, \bibinfo {author} {\bibfnamefont {S.~B.}\ \bibnamefont {Chung}},
  \bibinfo {author} {\bibfnamefont {C.-H.}\ \bibnamefont {Hsu}}, \bibinfo
  {author} {\bibfnamefont {I.}~\bibnamefont {Mandal}}, \bibinfo {author}
  {\bibfnamefont {S.}~\bibnamefont {Raghu}}, \ and\ \bibinfo {author}
  {\bibfnamefont {S.}~\bibnamefont {Chakravarty}},\ }\href {\doibase
  10.1103/PhysRevB.85.035110} {\bibfield  {journal} {\bibinfo  {journal} {Phys.
  Rev. B}\ }\textbf {\bibinfo {volume} {85}},\ \bibinfo {pages} {035110}
  (\bibinfo {year} {2012})}\BibitemShut {NoStop}%
\bibitem [{\citenamefont {Majumdar}\ and\ \citenamefont {Ghosh}(1969)}]{MG}%
  \BibitemOpen
  \bibfield  {author} {\bibinfo {author} {\bibfnamefont {C.~K.}\ \bibnamefont
  {Majumdar}}\ and\ \bibinfo {author} {\bibfnamefont {D.~K.}\ \bibnamefont
  {Ghosh}},\ }\href@noop {} {\bibfield  {journal} {\bibinfo  {journal} {J.
  Math. Phys.}\ }\textbf {\bibinfo {volume} {10}},\ \bibinfo {pages} {1399}
  (\bibinfo {year} {1969})}\BibitemShut {NoStop}%
\end{thebibliography}%


%

\end{document}